\documentclass[sn-nature,Numbered]{sn-jnl}
\usepackage{graphicx}%
\usepackage{multirow}%
\usepackage{amsmath,amssymb,amsfonts}%
\usepackage{amsthm}%
\usepackage{mathrsfs}%
\usepackage[title]{appendix}%
\usepackage{color}%
\usepackage{textcomp}%
\usepackage{manyfoot}%
\usepackage{booktabs}%
\usepackage{algorithm}%
\usepackage{algorithmicx}%
\usepackage{algpseudocode}%
\usepackage{listings}%

\usepackage{mathtools}
\usepackage{upgreek}
\usepackage[dvipsnames,table,xcdraw]{xcolor}
\usepackage{thmtools}
\usepackage{lineno}
\usepackage{dsfont}
\usepackage{hyperref}

\usepackage[normalem]{ulem}

\usepackage{outline}
\usepackage{physics}
\usepackage{siunitx}
\usepackage{tikz}
\usetikzlibrary{shapes.geometric, arrows,positioning}
\usepackage{qcircuit}
\usepackage{lipsum}
\usepackage{chemfig}
\usepackage{silence}
\usepackage{algorithm}
\usepackage{algpseudocode}

\renewcommand{\thesection}{\Roman{section}} 
\renewcommand{\thesubsection}{\thesection.\Roman{subsection}}

\newcommand{\Kieran}[1]{{\color{Black}#1}}

\newcommand{\Change}[1]{{\color{Black}#1}}

\newcommand{\pool}{\mathcal{P}}
\newcommand{\subpool}{\mathcal{S}}

\newcommand{\cnotset}{\mathcal{R}_{\mathrm{CX}}}

\newcommand{\pc}{p_{\textrm{c}}}
\newcommand{\ncx}{N_{\mathrm{II}}}

\newcommand{\circuit}{\mathcal{C}}

\def\approxprop{%
  \def\p{%
    \setbox0=\vbox{\hbox{$\propto$}}%
    \ht0=0.6ex \box0 }%
  \def\s{%
    \vbox{\hbox{$\sim$}}%
  }%
  \mathrel{\raisebox{0.7ex}{%
      \mbox{$\underset{\s}{\p}$}%
    }}%
}

\DeclareMathOperator*{\argmax}{argmax}
\DeclareMathOperator*{\argmin}{argmin}
\def\app#1#2{%
  \mathrel{%
    \setbox0=\hbox{$#1\sim$}%
    \setbox2=\hbox{%
      \rlap{\hbox{$#1\propto$}}%
      \lower1.1\ht0\box0%
    }%
    \raise0.25\ht2\box2%
  }%
}
\def\approxprop{\mathpalette\app\relax}

\begin{document}

\title{
  Quantifying the effect of gate errors on variational quantum eigensolvers for quantum chemistry
}

\author*[1,2,3]{\fnm{Kieran} \sur{Dalton}}\email{kd437@cantab.ac.uk}

\author[1,2]{Christopher K. Long}

\author[1,2]{Yordan S. Yordanov}

\author[1,2]{Charles G. Smith}

\author[2]{Crispin H. W. Barnes}

\author[1]{Normann Mertig}

\author[1]{David R. M. Arvidsson-Shukur}

\affil[1]{Hitachi Cambridge Laboratory, J. J. Thomson Ave., Cambridge, CB3 
0HE, United Kingdom}
\affil[2]{Cavendish Laboratory, Department of Physics, University of 
Cambridge, Cambridge, CB3 0HE, United Kingdom}
\affil[3]{Current Address: Department of Physics, ETH Z\"urich, CH-8093 Z\"urich, 
Switzerland}

\date{\today}

\abstract{\Change{
	Variational quantum eigensolvers (VQEs) are leading candidates to demonstrate near-term quantum advantage.
	Here, we conduct density-matrix simulations of leading gate-based VQEs for a range of molecules. We numerically quantify their level of tolerable depolarizing gate-errors.
	We find that:\
	(i) The best-performing VQEs require gate-error probabilities between 
	$10^{-6}$ and $10^{-4}$ ( $10^{-4}$ and $10^{-2}$ with error mitigation) to predict, within chemical accuracy, ground-state energies of small molecules with $4-14$ orbitals.
	(ii) ADAPT-VQEs that construct ansatz circuits iteratively outperform fixed-circuit VQEs.
	(iii) ADAPT-VQEs perform better with circuits constructed from 
	gate-efficient rather than physically-motivated elements.
	(iv) The maximally-allowed gate-error probability, $p_c$, for any VQE 
	to achieve chemical accuracy decreases with the number $\ncx$
	of noisy two-qubit gates as $p_c\approxprop\ncx^{-1}$. 
  Additionally, $p_c$ decreases with system size, even with error mitigation, 
  implying that larger molecules require even lower gate-errors. Thus, quantum 
  advantage via gate-based VQEs is unlikely unless gate-error probabilities are decreased by orders of magnitude.
}}

\maketitle

\section{Introduction}

Calculating the ground-state energy of a molecular Hamiltonian is an 
important but hard task in computational chemistry \cite{QCCReview}. 
For strongly correlated systems, exact classical approaches quickly become 
infeasible as system sizes exceed $100$ spin-orbitals. 
Other, approximate, methods often lack accuracy 
\cite{HF,KSDFT,KSelectrongas,FCI,QCCReview}.
This makes quantum computers an attractive alternative.
A potential route to quantum-chemistry simulations relies on the 
quantum-phase-estimation algorithm (QPEA) \cite{PhaseEstimation, QCCReview}. 
However, the QPEA requires executing millions of gates on 
error-corrected hardware \cite{QPEARiverlane}.
Realizing such hardware requires significant resource overheads and gate-error 
probabilities below a minimal threshold \cite{SurfaceCodeReview2}.
For example, the surface code \cite{SurfaceCode1, SurfaceCode2}, requires thousands of physical qubits to 
implement a single logical qubit at a gate-error probability of $10^{-4}$ 
\cite{SurfaceCodeReview}.
In view of these requirements, the QPEA is not yet feasible.

To reduce the qubit number and gate-error requirements, the variational quantum 
eigensolver (VQE) was  proposed \cite{VQEProposal}. The VQE is a hybrid quantum-classical algorithm that uses a classical optimizer and a parameterized quantum circuit, the `ansatz', to estimate ground-state energies.
Combined with significant hardware 
developments \cite{QCReviewGeneral}, VQEs have facilitated successful demonstrations of 
quantum computational chemistry for small systems \cite{VQEProposal, HFOnSCQC, ScalableQuantumSimulationOfMolecularEnergies, HardwareEfficient, 
UCCSDDemonstration, DelftLogic}. These demonstrations have been aided by VQE algorithms' abilities to correct for certain errors \cite{VQEResilienceTheory, 
ScalableQuantumSimulationOfMolecularEnergies, VQEResilienceAndQSE}. 
Despite these achievements, there are still significant hurdles to overcome for VQEs to become useful. First, the short, gate-efficient ansätze used in small-scale experimental demonstrations 
\cite{ScalableQuantumSimulationOfMolecularEnergies,HFOnSCQC,HardwareEfficient,
UCCSDDemonstration,DelftLogic,VQEProposal} face optimization difficulties for larger systems. 
This is related to the emergence of barren plateaus (vanishing gradients), which are more likely when the ansatz is unrelated to the Hamiltonian \cite{ADAPTBP, VQETraps}. Current research, for example on growing the ansatz circuit \textit{iteratively} (the ADAPT-VQE algorithm \cite{ADAPTVQE}) is aimed at avoiding or mitigating the issue of barren plateaus \cite{ADAPTBP}.
Another significant hurdle \Kieran{comes from} gate-error rates in  hardware. Although current noisy intermediate-scale quantum devices \cite{NISQ, IBMQuantumSystems, IBMNature} have sufficiently many qubits to run VQEs for molecules with more than 100 spin-orbitals \cite{QCReviewGeneral}, their gate-error rates are too high.

At present, efforts to ameliorate the gate-error issue aim to either reduce 
ansatz circuit depths \cite{ADAPTVQE, Qubit-ADAPT-VQE,YordanEfficientQuantumCircuits,YordanQEB, VQEReview} or implement elaborate error-mitigation schemes \cite{TemmeMitigation, Benjamin_2017,Benjamin_2021}. 
However, VQEs are often benchmarked in the absence of gate errors, 
with circuit depths and CNOT counts used as proxies of their noise resilience \cite{VQEReview}.
It has been argued that the maximum viable circuit depth 
for a VQE ansatz circuit is given by the reciprocal of gate-error probability $1/p$ \cite{QCCReview}. 
More rigorously, given a gate-error probability $p$, the maximum VQE circuit depth 
which cannot be simulated classically scales as $\mathcal{O}(p^{-1})$ 
\cite{VQELimitationsI, VQELimitations}.
A research question, which remains under-explored, is to quantify the gate-error 
probabilities that VQEs can tolerate.
Specifically, considering the analogy of a surface code, which has a 
well-defined fault-tolerance threshold \cite{SurfaceCodeReview}, we \Change{aim to find the maximally allowed gate-error probability, 
below which a certain VQE estimates a certain molecule’s energies within chemical accuracy.}
Quantifying the maximally allowed gate-error probability \Kieran{allows the noise resilience of leading VQEs to be ranked}, and provides useful goals for the hardware community.

In this article, we \Kieran{numerically} quantify under how high gate-error probabilities VQEs can operate successfully.
More specifically, using density-matrix simulations, we simulate the ground-state search of leading, gate-based VQEs for a range of molecules.
In the presence of depolarizing noise, we 
show that:\
(i) Even the best performing VQEs require gate error probabilities $\pc$ on the 
order of $10^{-6}$ to $10^{-4}$ (without error mitigation) in order to predict molecular ground-state 
energies within chemical accuracy of $1.6 \times 10^{-3}$ Hartree.
This is significantly below the fault-tolerance threshold of the surface code \cite{SurfaceCodeReview}.
For small systems, error mitigation can be employed such that the required $\pc$ values can be improved to $10^{-4}$ to $10^{-2}$.
(ii) ADAPT-VQEs tend to tolerate higher gate-error probabilities than VQEs 
that use fixed ansätze, such as UCCSD and k-UpCCGSD.
(iii) ADAPT-VQEs tolerate higher gate-error probabilities when circuits are 
synthesized from gate-efficient \cite{Qubit-ADAPT-VQE, YordanQEB,YordanEfficientQuantumCircuits, YordanExcited}, rather than physically-motivated \cite{ADAPTVQE}, elements.
We support these claims by estimating, in 
the presence of depolarizing noise, the scaling relation between the maximally tolerable 
gate-error probability $p_c$ and the number $\ncx$ of noisy (two-qubit) gates.
Our results indicate that $p_c\approxprop\ncx^{-1}$ for any gate-based VQE.
(iv) We find that the maximally allowed gate-error probability, $p_c$, decreases with system size, with and without error mitigation. 
This shows that larger molecules would likely require even lower gate-errors.
We conclude that substantial quantum advantage in VQE-based quantum chemistry 
is unlikely, unless gate-errors are significantly reduced, or error-corrected hardware is realized, or error-mitigation protocols are improved and made scalable.

\section{Results}

\subsection{ADAPT-VQEs}
\label{sec:VQESection}

In this work, we investigate several classes of VQE algorithms.
Our study prioritizes VQEs with short ansatz circuits, as these are expected to 
be more noise resilient \cite{YordanQEB, VQEReview}.
Specifically, we consider ADAPT-VQEs, which have comparatively short ansatz 
circuits \cite{ADAPTVQE,YordanQEB} and the ability to mitigate rough parameter landscapes \cite{ADAPTBP}.
We further consider UCCSD \cite{UCCSD} and k-UpCCGSD \cite{k-UpCCGSD} as prototypes of fixed ansatz VQEs -- 
the latter for its comparatively shallow ansatz circuits \cite{VQEReview}.
Before we outline the results of our noise-resilience investigation, we describe 
the workings of the (ADAPT-)VQE.

The main idea of VQEs is to use shallow ansatz circuits, defined by a set of 
parameters $\boldsymbol{\uptheta}$, to generate entangled trial states 
$\rho(\boldsymbol{\uptheta})$.
A classical optimizer is then used to vary $\boldsymbol{\uptheta}$ and minimize 
the energy-expectation value of  $H$.
Provided that the ansatz is sufficiently expressive, the Rayleigh-Ritz 
variational principle,
\begin{equation}
    \label{eq:RayleighRitz}
   E(\boldsymbol{\uptheta})=\Trace\left[H\rho(\boldsymbol{\uptheta})\right] \geq 
\mathcal{E}_{0},
\end{equation}
allows $\min_{\boldsymbol{\uptheta}} (E(\boldsymbol{\uptheta}))$ to approach the molecular 
ground-state energy $\mathcal{E}_{0}$ \cite{QCCReview}.
ADAPT-VQEs use a classical optimizer in two ways \cite{ADAPTVQE}: to conduct the 
Rayleigh-Ritz minimization with respect to a parameterized quantum state; and to 
iteratively construct the ansatz that generates the parameterized state itself. 
A quantum computer is used to calculate the energy-expectation value of the 
parameterized state.

Consider the state generated by the ansatz  $U_{n}$:
\begin{equation}
  \label{eq:TrialStates}
  \rho_{n}(\theta_1,\ldots,\theta_n) = 
U_{n}(\theta_1,\ldots,\theta_n)\rho_{0}\,U_{n}^{\dagger}(\theta_1,\ldots,
\theta_n) .
\end{equation}
 $ \rho_{n}(\theta_1,\ldots,\theta_n) $ is parameterized by $n$ parameters. In ADAPT-VQE, the classical 
optimizer and the quantum computer work to find a minimum-energy expectation 
value:
\begin{equation}
  \label{eq:EnergyExpectation}
E_n \equiv \min_{\theta_1,\ldots,\theta_n}
\Trace\left[H\rho_{n}(\theta_1,\ldots, \theta_n)\right].
\end{equation}
An ADAPT-VQE iteratively adds parameterized elements to its ansatz to construct 
$\rho_{n}(\theta_1,\ldots,\theta_n) $ such that $E_1 > \ldots > E_n$ and $E_n$ 
approaches  $\mathcal{E}_0$.

The iterative ansatz construction proceeds as follows.  First,  the ADAPT-VQE 
algorithm initializes a state $\rho_{0}$, usually the Hartree-Fock state 
\cite{HF}. Then, the algorithm generates a sequence of  trial states by 
successively adding  elements of the form
\begin{linenomath} \begin{align}
    \label{eq:AnsatzElement}
    A_{\alpha}(\theta) = e^{\theta T_{\alpha}},
\end{align} \end{linenomath}
picked from a finite pool $\pool$ of operators (see below).
Here, $T_{\alpha}$, for $\alpha \in [1,\ldots, |\pool|]$, 
are anti-Hermitian operators.
Thus, the unitary ansatz grows as
\begin{linenomath} \begin{align}
    U_0 &= \mathds{I} , \\
    U_n(\theta_{1},\ldots,\theta_{n}) &= 
A_n(\theta_n)U_{n-1}(\theta_{1},\ldots,\theta_{n-1}).
\end{align} \end{linenomath}
The ansatz element $A_{n}(\theta_n)\in\pool$ is typically picked to yield the 
steepest energy gradient.
For each value of $\alpha$, a quantum computer evaluates the energy expectation 
value after adding $A_{\alpha}(\theta_n)$ in the $n$th step:
\begin{equation}
  \label{eq:LocalEnergyExpectation}
E_{n,\alpha}(\theta_n) \equiv \Trace\left[ H A_{\alpha}(\theta_n) \rho_{n-1} 
                                      A_{\alpha}^{\dagger}(\theta_n)\right] .
\end{equation}
The ADAPT-VQE algorithm then picks the element $A_n \equiv 
A_{\alpha=\alpha_n}$ with 
\begin{linenomath} \begin{align}
  \label{eq:Gradient}
  \alpha_n &= \argmax_{\alpha: A_\alpha\in\mathcal P}
  \left|\left.\frac{\partial E_{n,\alpha}(\theta_n)}{
                    \partial \theta_n}\right|_{\theta_n=0}\right| \\
&= \argmax_{\alpha: A_\alpha\in\mathcal P} \left|\Trace\left\{\left[H, 
T_\alpha\right]\rho_{n-1}\right\}\right|. \nonumber
\end{align} \end{linenomath}
Alternatively, one may define a sub-pool $\subpool\subset\pool$ of operators 
with the largest gradients and let the algorithm pick the element with the 
largest energy difference:
\begin{equation}
    \label{eq:EnergyDifference}
  \alpha_n = \argmin_{\alpha: A_\alpha\in\subpool} 
\left[\underset{\theta_n}{\min}E_{n,\alpha}(\theta_n)\right] .
\end{equation}

After choosing the $n$th ansatz element $A_{n}(\theta_n)$, a classical computer
optimizes and updates the parameters $\theta_1, \ldots, \theta_n$ to minimize 
the energy expectation value $E_n$ in Eq.~\eqref{eq:EnergyExpectation}.
Provided that   $E_n - E_{n-1} > \epsilon$, for some energy precision $\epsilon$,  the iterative algorithm continues. 
When $E_n - E_{n-1} \leq \epsilon$ the algorithm halts at some final length $n=N$, and outputs $E_N \equiv 
E_n$ as the estimate of $\mathcal{E}_0$.

In this work, we focus on the three main types of ADAPT-VQEs: 
fermionic-ADAPT-VQE, QEB-ADAPT-VQE and qubit-ADAPT-VQE. (Efficient gate-representations for their relevant ansatz elements can be 
found in \cite{YordanEfficientQuantumCircuits, YordanQEB}.)
These algorithms differ in their ansatz-element pools $\pool$.

First, we consider the fermionic-ADAPT-VQE \cite{ADAPTVQE}. As the name suggests, 
this algorithm uses a pool of operators that closely simulate the 
physics of fermionic excitations.
The pool is formed from
\begin{linenomath} \begin{align}
  \label{eq:fermionic}
  T_{i k} &= a_{i}^{\dagger}a_{k}-a_{k}^{\dagger} a_{i}, \; \; \mathrm{and} \\
  T_{i j k l} &= a_{i}^{\dagger} a_{j}^{\dagger} a_{k} a_{l}
                -a_{k}^{\dagger} a_{l}^{\dagger} a_{i} a_{j}.
\end{align} \end{linenomath}
Here, $a_{i}^{\dagger}$ and $ a_{i}$ are fermionic creation and annihilation 
operators acting on the $i$th orbital. 
Throughout this work,  we represent these operators using  the Jordan-Wigner 
transformation \cite{JordanWigner}, where
\begin{linenomath} \begin{align}
a_{i}^{\dagger}&\mapsto \frac{1}{2}\left(X_{i}-\mathrm{i} Y_{i}\right) 
\prod_{r=0}^{i-1} Z_{r}, \; \; \mathrm{and}  \\
a_{i}&\mapsto \frac{1}{2}\left(X_{i}+\mathrm{i} Y_{i}\right) \prod_{r=0}^{i-1} 
Z_{r}.
\end{align} \end{linenomath}
$X_i, Y_i, Z_i$ are the Pauli operators acting on the $i$th qubit.
The fermionic-ADAPT-VQE leads to shallower and more gate efficient circuits than UCCSD \cite{ADAPTVQE}.
Further, choosing fermionic excitations along the gradient of minimum energy 
produces $H$-tailored circuits, which can potentially avoid barren plateaus 
\cite{ADAPTBP}.

Second, we  consider the QEB-ADAPT-VQE \cite{YordanQEB,YordanExcited}. 
This algorithm uses a pool of operators that nearly (up to a $\pm$ sign) simulate 
the physics of fermionic excitations.
The pool is formed from
\begin{linenomath} \begin{align}\label{eq8}
 T_{i k}&=  Q_{i}^{\dagger} Q_{k}-Q_{k}^{\dagger} Q_{i}, \; \; \mathrm{and} \\
 T_{i j k l}&= Q_{i}^{\dagger} Q_{j}^{\dagger} Q_{k}Q_{l}
                     - Q_{k}^{\dagger} Q_{l}^{\dagger} Q_{i}Q_{j},
\end{align} \end{linenomath}
where
\begin{linenomath} \begin{align}
 Q_{i}^{\dagger} \coloneqq \frac{1}{2}\left(X_{i}-\mathrm{i} Y_{i}\right)
 \text{ and } Q_{i} \coloneqq \frac{1}{2}\left(X_{i}+\mathrm{i} Y_{i}\right).
\end{align} \end{linenomath}
$Q_{i}^{\dagger}$ and $ Q_{i}$ are known as qubit creation and annihilation 
operators, respectively.
Due to the CNOT efficiency of its pool, the QEB-ADAPT-VQE  can find ground-state 
and excited-state energies with fewer CNOT gates than the
fermionic-ADAPT-VQE \cite{YordanEfficientQuantumCircuits, YordanQEB, 
YordanExcited}. 

Finally, we  consider the qubit-ADAPT-VQE \cite{Qubit-ADAPT-VQE}. 
This algorithm uses a pool of gate-efficient elements without 
physical motivation.
The pool is formed from segments of Pauli-operator strings:
\begin{linenomath} \begin{align}
    \label{eq:Pauli}
    T_{ij} &= i \sigma_i \sigma_j , \; \; \mathrm{and}  \\
    T_{ijkl} &= i \sigma_i \sigma_j \sigma_k \sigma_l ,
\end{align} \end{linenomath}
where $\sigma_{i}$ denotes Pauli operators $X_i, Y_i, Z_i$ acting on the $i$th qubit.
In previous works, this pool has been found to generate the most shallow and 
CNOT efficient circuits for ADAPT-VQEs \cite{Qubit-ADAPT-VQE}. In our simulations, 
we use a pool formed from $XY$-Pauli strings of length two and four with an odd number of $Y$'s. 
It is possible to use reduced pools \cite{Qubit-ADAPT-VQE}, but at the expense of reduced  circuit efficiency 
of the final ansatz \cite{YordanQEB}.

Typically, the fermionic-ADAPT-VQE and the qubit-ADAPT-VQE use the gradient-based decision rule expressed in Eq.~8. 
On the other hand, the original QEB-ADAPT-VQE uses the energy-based decision rule, shown in Eq.~9. These algorithms are 
summarized in a flow-chart summary in Fig.~\ref{fig:flowchart}, and in the pseudocode of Supplementary Note~1.


\begin{figure}
\includegraphics[width=\columnwidth]{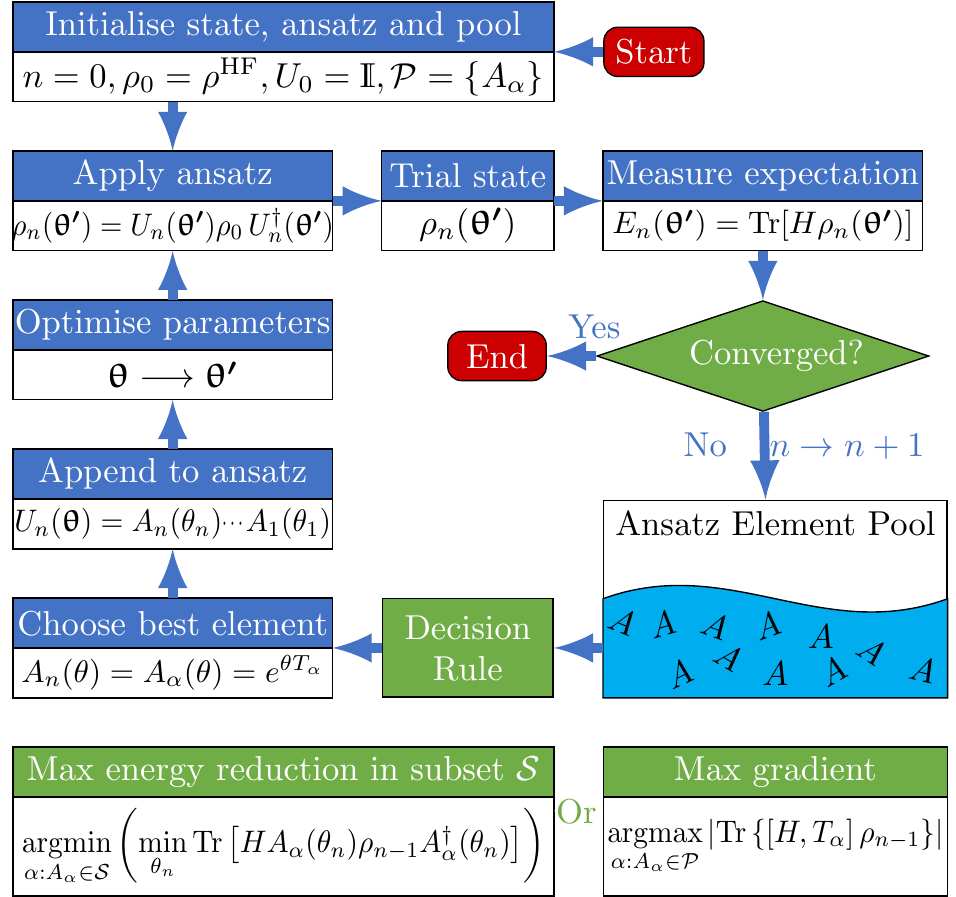}
    \caption{Flowchart of the ADAPT-VQE algorithm.\label{fig:flowchart}}
\end{figure}

To demonstrate the benefits of iteratively-grown ansätze, we compare them to 
a typical fixed-ansatz-VQE method: the UCCSD-VQE \cite{UCCSD, UCCSDDemonstration}. 
In Supplementary Note~4, we extend this comparison to the k-UpCCGSD algorithm \cite{k-UpCCGSD}. Owing to 
its linear scaling of circuit depth with qubit number, this algorithm was recently put forward as the leading 
fixed-ansatz VQE \cite{VQEReview}.
We simulate the workings of the fixed-ansatz methods using the aforementioned fermionic and QEB elements.

Given the breadth of work on VQEs \cite{VQEReview}, it is not possible to perform an exhaustive analysis 
of all existing algorithms. Nevertheless, the analytical results in Sec.~\ref{sec:scaling}, and the low circuit depths provided by ADAPT-VQE \cite{VQEReview}, suggest that our 
results provide a lower bound on the requirements for gate-based VQE algorithms to operate successfully. However, there exist 
algorithms that differ greatly from typical VQEs, and could deserve future attention. 
We discuss some of these, and the reasons for our exclusion of them, below.
We will not consider iterative qubit coupled cluster (iQCC) \cite{iQCC} and ClusterVQE 
\cite{ClusterVQE} algorithms. 
We do not anticipate these algorithms to be feasible options to study strongly 
correlated systems, whose simulation using quantum algorithms provides the most 
benefit over classical algorithms.
We also omit the  DISCO-VQE \cite{Disco-VQE}.
Due to its large jumps in Hilbert space during the discrete optimizations 
of the ansatz, we expect DISCO-VQE to lack tolerance to barren plateaus.
These problems may be overcome in future improvements of these VQE algorithms.
We leave the design of improved algorithms, and the noise-evaluation of them, 
to future articles.
Finally, we omit the  ctrl-VQE algorithm \cite{Ctrl-VQE}. Although highly interesting, this Hamiltonian algorithm operates 
with device-tailored pulses, rather than quantum gates, and thus lies outside 
the scope of this work.

\subsection{Density-matrix simulations}
\label{sec:densitymatrixsimulations}

To investigate the effect of noise on gate-based VQE, 
we constructed a VQE-tailored density-matrix simulator, expanding the 
state-vector circuit simulator of Ref.~\cite{YordanQEB}.
We represent molecular orbitals in the Slater type orbital-3 Gaussians (STO-3G) 
spin-orbital basis set \cite{STO-3G}, with the option of frozen orbitals.
The openfermion-Psi4 package \cite{OpenFermion, Psi4} is used to generate the 
second-quantized Hamiltonian and to perform the Jordan-Wigner transformation 
\cite{JordanWigner}.
Ansatz parameters  are optimized using Nelder-Mead 
\cite{Nelder-Mead} or gradient-descent-based (BFGS) \cite{BFGS} methods in SciPy 
\cite{SciPy}.

We note that, due to the wide array of quantum-computing platforms and their 
contrasting qubit-control implementations, no noise model can be simultaneously realistic and platform-agnostic. 
In this work, we model noise by applying single-qubit depolarizing noise 
to the target qubit $i$ after the application of each two-qubit CNOT gate. 
Our noise channel can be represented by
\begin{equation}
    \mathcal 
D(i,p)\left[\rho\right] \coloneqq 
\left(1-p\right)\rho+\frac{p}{3}\sum_{\sigma_{i}}
\sigma_i\rho\sigma_i,
  \label{eq10}
\end{equation}   
where $p \in [0,1]$ is the gate-error probability.

In real devices, noise from two-qubit gates completely dominates the noise from 
single-qubit gates \cite{SQCReview,IBMQuantumSystems,Ion1,Ion2}. Thus, we ignore the latter.
Additionally, we exclude state preparation and measurement errors, which are 
often lower in magnitude than the accumulated two-qubit gate errors 
\cite{SQCReview}, and can be mitigated efficiently in experiments \cite{SPAM1, 
SPAM2, SPAM3, SPAM4, McWeeny, McWeenyBenchmark}. \Kieran{(We note that ADAPT-VQE algorithms have 
high measurement requirements, such that measurement errors may prevent the algorithm from reaching 
the global minimum energy. This topic requires further investigation.)}
Depolarizing noise is commonly used to represent local and Markovian gate errors 
when assessing both NISQ \cite{NISQDep1,NISQDep2,NISQDep3} and 
quantum-error-correction \cite{SurfaceCodeWithDecoherence, QECDep2, QECDep3, 
SurfaceCodeReview} algorithms.
More realistic models can include thermal-relaxation noise (dephasing and 
amplitude damping) \cite{ThermalRelaxationNoise} and device-specific 
gate errors derived from gate-set-tomography data \cite{GST}.
When $T_1\approx T_2$ thermal relaxation noise can be 
approximated using our depolarizing noise model 
\cite{SurfaceCodeWithDecoherence}.
This is a reasonable model for superconducting hardware \cite{SQCReview}. 
On the other hand, when $T_2\ll T_1$ dephasing noise dominates our depolarizing noise model is less accurate. 
This is common in trapped-ion devices \cite{Ion1, Ion2} and spin qubits \cite{SpinReview} and has recently been investigated in Ref.~\cite{ChrisPaper}. 
Moreover, existing VQE algorithms require unrealistically low error rates to give chemically accurate energies. Any attempt to scale down the error rates in realistic noise models to these low levels must be theoretically-justified. This is challenging for a complex, multi-parameter model.
Hence, we exclude noise models based on gate-set tomography.
Finally, we do not consider coherent errors, since their effect can 
be suppressed by randomized compiling \cite{RandomizedCompiling} and dynamical 
decoupling \cite{DynamicalDecouplingA, DynamicalDecouplingB}.
Randomized compiling \cite{RandomizedCompilingConversion} can also be 
used to convert coherent errors to stochastic errors.
Additionally, VQE algorithms are somewhat resilient to coherent errors 
\cite{VQEProposal, BiamonteCoherentError}. Thus, in this work we focus on 
incoherent errors. \Kieran{Note that, if VQE algorithms were studied with a coherent noise model, their perceived performance may be greater.}

When simulating the smallest molecules ($\mathrm{H}_2$ and $\mathrm{H}_4$) we 
apply our noise channel [Eq. \eqref{eq10}] after each application of a CNOT 
gate.  This gate-by-gate method is computationally expensive.
To facilitate feasible simulations of molecules larger than  $\mathrm{H}_4$, we 
approximate 
each noisy ansatz element by a corresponding noiseless ansatz evolution and a 
noise-inducing evolution. The noise-inducing evolution corresponds to 
depolarizing noise applied to each qubit in accordance with the number of times 
that qubit was a CNOT target in the ansatz element. We observe that this 
lower-bounds the effect of noise. For example, for $\mathrm{H}_4$, applying 
noise after each CNOT with gate-error probability $p = p^{\prime}$, gives 
approximately the same 
energy accuracy as applying total noise after each element with 
$p \approx 1.3 p^{\prime}$. Consequently, our simulations of larger molecules 
should not be compared directly with those for $\mathrm{H}_2$ and 
$\mathrm{H}_4$. A detailed illustration of our noise approximation is given in Supplementary Note~3.

Energy accuracy is the key metric of VQE 
performance.
It is defined as
\begin{linenomath} \begin{align}
    \Delta E(p,n) \coloneqq E_n(p) - E_{\mathrm{FCI}}.
\end{align} \end{linenomath}
Here, $E_n(p)$ is the VQE-calculated  energy with gate-error probability $p$ in 
the $n$th iteration and $E_{\mathrm{FCI}}$ is the energy given by the full-configuration-interaction 
\cite{FCI} calculation of the true ground-state energy $\mathcal{E}_0$.
A key objective of our study is to find the maximally allowed gate 
error probability $p_c$ for which $\Delta E(p,n)<1.6$ milli-Hartree.

Classical optimizers are used to 
tune $\theta_1,\ldots,\theta_n$. The parameters are optimized until the 
gradient norm, $\left|\nabla_{\vec\theta}E\right|\le\epsilon_O$, for some 
precision $\epsilon_O$.  In our simulations of $\mathrm{H}_2$ we calculate 
converged values of $E_n(p)$ using our density-matrix simulator. To keep 
larger-molecule simulations tractable, we estimate $\Delta E$ as follows.
We first grow the ansatz circuit $\mathcal{C}_n$ in noiseless, unitary 
simulations until the $n$th iteration, for which the energy accuracy $\Delta 
E(0,n)$ first drops below a cut-off energy precision $\epsilon_t$: $\Delta 
E(0,n)<\epsilon_t$.
Then, we approximate $\Delta E(p,n)$ by simulating the implementation of  
$\mathcal{C}_n$ with noise on our density-matrix simulator.
Thus, $\Delta E(p,n)$ may depend on the iteration $n$. 
As demonstrated in Supplementary Note~2, ansatz growth and optimization in the presence of noise has little effect on the noise probability required for chemical accuracy.
%
%

\begin{figure*}[]
  \includegraphics[width=\textwidth,trim=5 3 7 3, 
  clip]{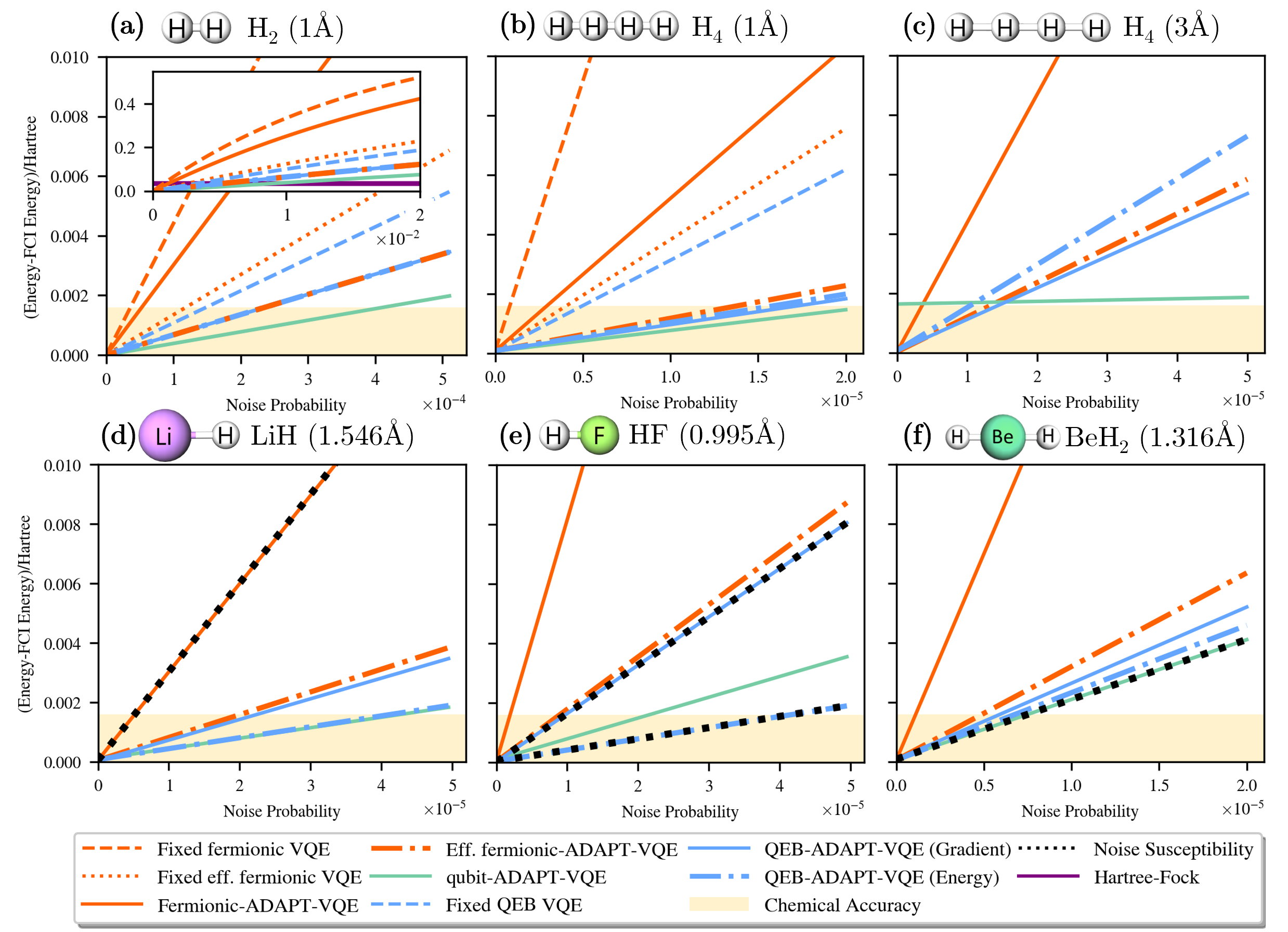}
  \caption{Energy accuracy as a function of gate-error probability for 
$\mathrm{H}_{2}$, 
  $\mathrm{H}_4$ (at 1\AA\ and 3\AA\ interatomic separation), $\mathrm{LiH}$, 
  $\mathrm{HF}$ and $\mathrm{BeH}_{2}$. Ansätze using fermionic, qubit, and 
  Pauli string elements are 
  plotted in red, blue and green, respectively. All curves labelled as ``fixed'' 
  VQE ansätze use the UCCSD ansatz \cite{UCCSD,UCCSDDemonstration}. Energy accuracies 
  lower than chemical accuracy are highlighted by the yellow region. The purple 
  line in the $\mathrm{H}_{2}$ inset is the energy calculated using the 
  Hartree-Fock \cite{HF} state. 
  Extrapolated noise-susceptibility calculations are 
  shown in black for the fermionic-ADAPT-VQE (\textbf{d}), the QEB-ADAPT-VQE 
(\textbf{e}) and 
  the qubit-ADAPT-VQE (\textbf{f}).\label{fig:1DPlots}}
\end{figure*}

\subsection{Comparison between ADAPT-VQEs with noise}
\label{1DPlotsSection}

In this section,  we benchmark the noise resilience of ADAPT-VQEs using our 
density-matrix simulator. We study $\mathrm{H}_2$, $\mathrm{H}_4$, 
$\mathrm{LiH}$, 
$\mathrm{HF}$ and $\mathrm{BeH}_{2}$. 
Our simulations were conducted using a parameter optimization cut-off of $\left|\nabla_{\vec\theta}E\right|\le\epsilon_O=10^{-6}$ 
Hartree and ansatz growth cut-off of $E_n - E_{n-1} \leq \epsilon=10^{-12}$ Hartree.
In our simulations of the larger molecules, we used an ansatz-truncation cut-off 
of $\Delta E(0,n)<\epsilon_t=10^{-4}$ Hartree. Below, we use $\Delta E(p) = \Delta E(p,n_{\mathrm{final}})$ 
to refer to the energy accuracy at the final ansatz length $n=n_{\mathrm{final}}$. Because of the significant skepticism towards error-mitigation strategies \cite{VQELimitations,EMScaling,EMScaling2}, we omit such strategies from the analyses presented in this section and investigate error mitigation separably in Sec.~II F.

\refstepcounter{table}\label{1DErrorTable}

The inset of Fig. \ref{fig:1DPlots}(a) shows how $\Delta E(p)$ varies with $p$ for 
$\mathrm{H}_2$.  The values of $p\in [0, 0.02]$ include the well-known surface-code fault-tolerance 
threshold \cite{SurfaceCodeReview, 
SurfaceCodeReview2} as well as the gate-error probability of currently available 
quantum hardware \cite{SQCReview,IBMQuantumSystems,Ion1,Ion2}.  All tested VQE algorithms require extremely small gate-error 
probabilities if they are to improve on the Hartree-Fock energy approximation, even for the 
simple $\mathrm{H}_2$ molecule. The region of chemical accuracy is too small to show in the inset. 
In real implementations of ADAPT-VQE algorithms, energies exceeding the Hartree-Fock approximation 
would not be achieved, since, in this case, adding elements to the ansatz does not improve the initial 
energy accuracy. Here, we add noise to a noiselessly-grown ansatz  such that these energies are shown. 
These observations motivated us to reduce significantly the range of $p$ used in the rest of this study.

The rest of Fig.~\ref{fig:1DPlots} shows our calculations of $\Delta E(p)$ 
as a function of $p$ for all considered molecules.  The region of 
chemical accuracy is highlighted by yellow shading.  We emphasize six general 
trends supported by our data.
First, the maximally-allowed gate-error probabilities $p_{\mathrm{c}}$ 
for computing  ground-state energies within chemical accuracy are extremely 
small. For all molecules investigated in this study, the value of 
$p_{\mathrm{c}}$ is on the order of $10^{-6}$ to $10^{-4}$ (see Table 
\ref{1DErrorTable} for details). These values are significantly below the 
fault-tolerance thresholds of leading error-correction protocols.
Second, our simulations of $\mathrm{H}_2$ and $\mathrm{H}_4$ (1Å) 
suggest that ADAPT-VQEs outperform fixed ansatz methods. For a given pool of 
ansatz elements, the corresponding ADAPT-VQE algorithm leads to better energy 
accuracies than the corresponding fixed ansatz VQE algorithm. 
Third, the efficient representation of fermionic excitations 
\cite{YordanEfficientQuantumCircuits} improves the performance of the 
fermionic-ADAPT-VQE significantly. This representation reduces CNOT depth, but 
its scaling of CNOT depth with molecule size is still worse than the scaling of 
QEB and Pauli string elements. The second and third observations 
support the claim \cite{YordanQEB} that the CNOT count is a useful estimator of VQE's 
noise vulnerability.
Fourth, the more gate-efficient  (Pauli string and QEB) pools 
outperform the most physically-motivated (fermionic) pool. The 
fermionic-ADAPT-VQE is consistently outperformed by either the qubit-ADAPT-VQE 
or the QEB-ADAPT-VQE.
Fifth, sometimes the QEB-ADAPT-VQE outperforms the qubit-ADAPT-VQE 
and \textit{vice versa}. For $\mathrm{H}_2$, $\mathrm{H}_4$ (1Å) and 
$\mathrm{BeH}_{2}$, the qubit-ADAPT-VQE outperforms the QEB-ADAPT-VQE.  On the 
other hand, for $\mathrm{HF}$, the QEB-ADAPT-VQE (energy-based decision rule) 
outperforms the qubit-ADAPT-VQE. For $\mathrm{LiH}$, the QEB-ADAPT-VQE and the 
qubit-ADAPT-VQE perform similarly. Notably, for $\mathrm{H}_4$ (3\AA), the 
qubit-ADAPT-VQE fails to add more than two elements to the ansatz. Hence, it 
never surpasses chemical accuracy. This gives some indications of the 
qubit-ADAPT-VQE's being  worse than the QEB-APAPT-VQE at 
simulating strongly correlated molecules.
Sixth, different decision rules for QEB-ADAPT-VQEs yield different  
performances.  For  $\mathrm{HF}$, $\mathrm{LiH}$ and $\mathrm{BeH}_{2}$, the  
energy-reduction decision rule gives a better energy accuracy than the 
maximum-gradient rule. Conversely, for $\mathrm{H}_{4}$ (1{\AA} and 3{\AA}) the 
gradient-based decision rule performs better. A study of the optimal decision 
rules for various molecular-energy landscapes is left for the future.

We close this subsection with a comment on benchmarks of fixed-ansatz 
VQEs. Both UCCSD and k-UpCCGSD ansätze have been investigated with numerical simulations. 
However, their energy accuracies significantly worse than those obtained using  
ADAPT-VQEs. In particular, the energy accuracies for the k-UpCCGSD 
algorithm do not fit the scale of Fig.~\ref{fig:1DPlots}. Hence, these 
results are presented separately, in Supplementary Note~4.

\subsection{Optimal truncation of iteratively-grown ansätze with noise}
\label{3DPlotsSection}

\begin{figure*}[]
\hspace*{-1cm}
\includegraphics[width=1.1\textwidth, trim=5 0 7 0, 
clip]{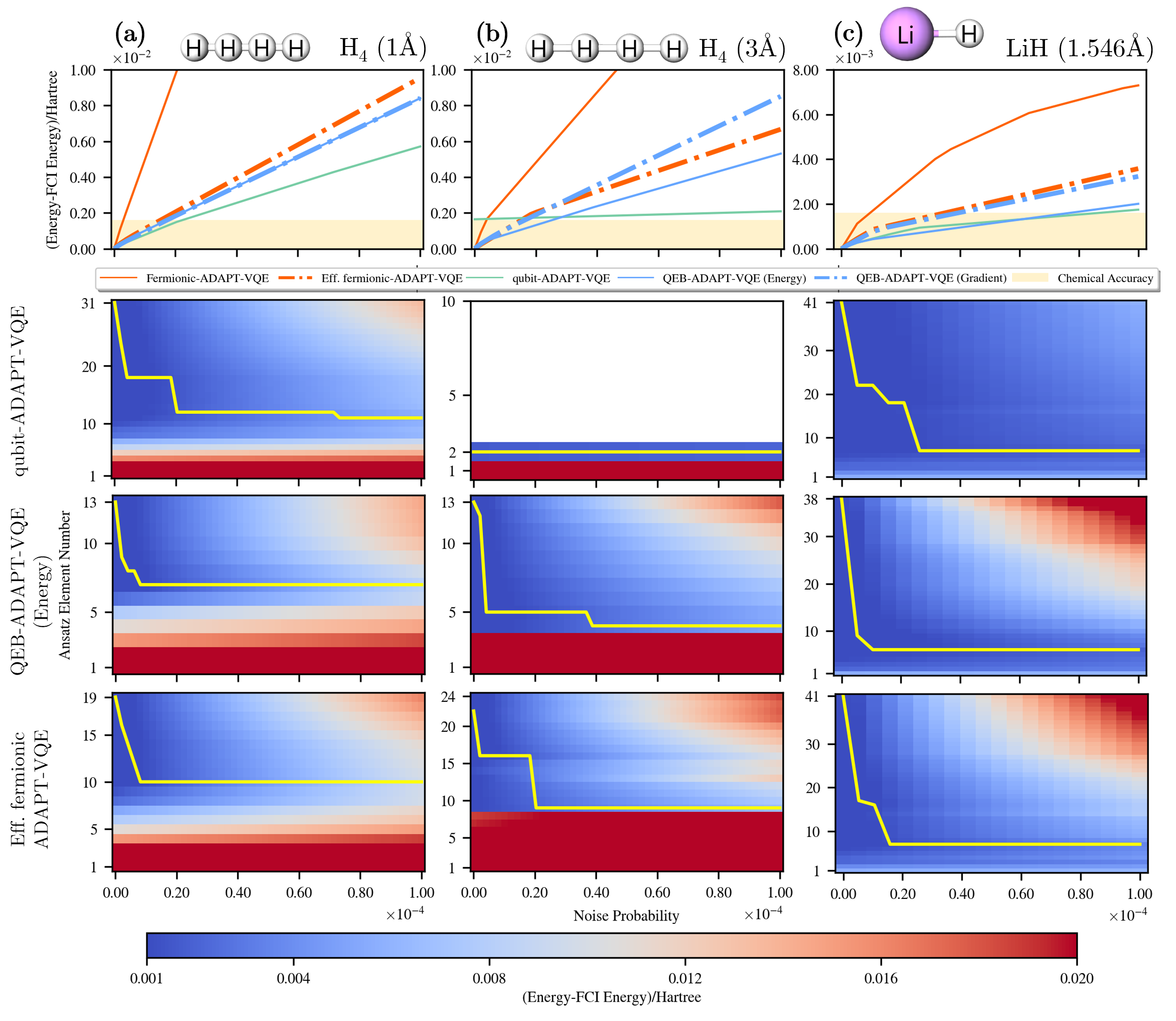}
\caption{
Colour plots representing the energy accuracy (from FCI) at different 
gate-error probabilities and ansatz lengths, for three different molecules. Three 
iterative-growth methods are included: the fermionic-ADAPT-VQE with efficient 
elements, the QEB-ADAPT-VQE with the energy decision rule and the qubit-ADAPT-VQE (the 
plots for the remaining two methods are given in Supplementary Note~2). The yellow lines on the colour plot highlight the ansatz lengths 
that minimize the energy accuracies for each gate-error probability. The 
top-column figures are 
extracted from the colour plots by plotting the energy accuracy along these curves. \label{fig:3DPlots}}
\end{figure*}


In the above analyses of ADAPT-VQEs for larger molecules, we  truncated 
noiselessly-grown ansätze in the $n$th iteration when an energy precision of 
$\epsilon_t$ was reached. Thus, we established a performance hierarchy between 
different VQEs. 
However, deeper circuits are generally more vulnerable to noise, and fixing $n$ in 
noiseless simulations could generate artificially deep circuits. Therefore, to 
showcase the full benefit of ADAPT-VQEs in the presence of noise, one must vary the truncation 
length. This would happen automatically for an ansatz grown with noise, as any 
truncation criterion would be met with shorter ansätze.
Here, we simulate this to conduct an alternative, more exact and more computationally expensive, 
comparison between ADAPT-VQEs. We optimize energy accuracy with respect to 
ansatz length:
\begin{linenomath} \begin{align}
 \label{eq:minimalenergyaccuracy}
 \Delta E(p, n_{\mathrm{opt}}) \equiv  \min_{n}\left\{\Delta E(p, n)\right\}.
\end{align} \end{linenomath}
Below, we present numerical results for $\mathrm{H}_4$ (1Å),  $\mathrm{H}_4$ 
(3Å) and $\mathrm{LiH}$ 
in the left, middle and right columns of Fig.~\ref{fig:3DPlots}, 
respectively. The colour plots show ADAPT-VQE simulations  of $\Delta E(p,n)$ as 
a function of $n$ and $p$.  The top row of Fig.~\ref{fig:3DPlots} shows  the 
optimized $\Delta E(p, n_{\mathrm{opt}})$ as a function of $p$. \Kieran{Additional colour plots for 
the VQE methods not shown in Fig.~\ref{fig:3DPlots} are provided in 
Supplementary Note~5.}

The data in Fig.~\ref{fig:3DPlots} warrant four comments.  
First,  in the absence of noise ($p=0$),  the energy accuracy decreases 
monotonically as the ansatz length ($n$)
increases, eventually surpassing chemical accuracy.  This is expected from the 
nature of the ADAPT-VQE methods.  The exception is the qubit-ADAPT-VQE simulation of $\mathrm{H}_4$ (3Å), which simply fails 
after $n=2$ and does not achieve chemical accuracy.
Second,  at small but finite values of $p$, $\Delta E(p,n)$ initially decreases 
with larger $n$. However, after an optimal length $n_{\mathrm{opt}}(p)$, the 
improvement from appending additional ansatz elements is outweighed by the 
detrimental increase of noise.  At this point,   $\Delta E(p,n)$ starts to 
increase 
with $n$. Thus, the ideal ansatz truncation happens at $n = n_{\mathrm{opt}}(p)$.  The $n_{\mathrm{opt}}(p)$ 
values are shown as yellow lines on the colour plots of 
Fig.~\ref{fig:3DPlots}.
These curves show $n_{\mathrm{opt}}(p)$ decreasing monotonically as $p$ increases.
Third,  plots of $ \Delta E(p, n_{\mathrm{opt}}) $ as a function of $p$ (top 
Fig.~\ref{fig:3DPlots}), show the same relative performance between 
the fermionic-ADAPT-VQE, the QEB-ADAPT-VQE and the qubit-ADAPT-VQE as in 
Fig.~\ref{fig:1DPlots}. 
Fourth,  the values for $p_{\mathrm{c}}$ increase when the ansätze are truncated 
at an optimal value of $n$, as compared to the arbitrary truncation used when 
producing Fig.~\ref{fig:1DPlots}. The values of $p_{\mathrm{c}}$ are presented 
in Table \ref{1DErrorTable} and Fig.~\ref{fig:Table1}.
After this optimal-truncation analysis, our overall conclusion remains unchanged:\ Even for the 
best-performing ADAPT-VQEs, the maximally allowed gate-error probability is on 
the order $10^{-6}$ to $10^{-4}$ Hartree.

\subsection{Analytical noise-susceptibility analysis}

\label{sec:scaling}

To analytically support our numerical results, we study the linear response of energy accuracy $\Delta E(p)$ to noisy perturbations of the 
unitary ansatz circuits.
Then, we use our  results to show that  $\pc$ is roughly inversely proportional to the number $N_{\mathrm{II}}$ of noisy 
(two-qubit) gates. 

\textit{Noise susceptibility:---}From Fig.\ref{fig:1DPlots} we see that  $\Delta E(p) \approx \chi^{\prime} p$, 
for some constant $\chi^{\prime}$. Inspired by this observation, we define the 
\textit{noise-susceptibility parameter}:
\begin{linenomath} \begin{align}
 \chi = \left.\frac{\partial \Delta E(p)}{\partial p}\right|_{p=0}.
\end{align} \end{linenomath}
Now, we show that $ \chi \propto \ncx$ (details are given in 
Supplementary Note~6).
If $p=0$, an ansatz circuit $\circuit$ can be expressed as a product of $R$ unitary gates: $U = G_R \cdots G_1$.
We use $\cnotset$ to denote the set of indices for which $G_r$ is a 
noisy (CNOT) gate, and we use $i_r$ to denote the qubit which noise acts on. 
Further,  we define a perturbed version of the target unitary $U$ as 
\begin{linenomath} \begin{align}
 U_{\mathrm{p}}(\sigma, r,{i_r}) &= G_R\cdots G_{r+1}\sigma_{i_r}G_{r}\cdots G_{1},
\end{align} \end{linenomath}
 where the Pauli gate $\sigma$ acts on qubit $i_r$ after the $r$th 
gate.
The corresponding energy expectation values are
\begin{subequations}
\label{eq:SusceptibilityEnergyExpectations}
\begin{linenomath} \begin{align}
E_{U}                   &= \Trace\left[H U \rho_{0} U^{\dagger}\right],\\
E_{ U_{\mathrm{p}}}(\sigma, r, {i_r}) &= 
\Trace\left[H U_{\mathrm{p}} (\sigma,r,{i_r}) \rho_{0}U_{\mathrm{p}}^{\dagger}(\sigma, r,{i_r}) \right].
\end{align} \end{linenomath}
\end{subequations}
Usually, $E_{U}$ is close to $\mathcal{E}_0$. Thus, we interpret 
$E_{ U_{\mathrm{p}}}(\sigma, r, {i_r})$ 
as a noise-induced excitation. We call $E_{ U_{\mathrm{p}}}(\sigma, r, {i_r}) - E_{U}$ the 
noise-induced 
fluctuation.
The average noise-induced fluctuation of the ansatz is
\begin{linenomath} \begin{align}
 \label{eq:AverageNoiseFluctuation}
\delta E  \equiv \frac{1}{\ncx} 
\sum_{r\in\cnotset} 
\frac{1}{3} \sum_{\sigma\in \{X,Y,Z\}} \left[E_{ U_{\mathrm{p}}}(\sigma, r, i_r) - E_U \right],
\end{align} \end{linenomath}
In Supplementary Note~6, we show that 
\begin{linenomath} \begin{align}
 \label{eq:NoiseSusceptibility}
 \chi =  \delta E \times \ncx.
\end{align} \end{linenomath}
Below, we analyze this expression for the noise-susceptibility parameter.

\begin{figure}
\centering
\includegraphics[width=0.6\columnwidth]{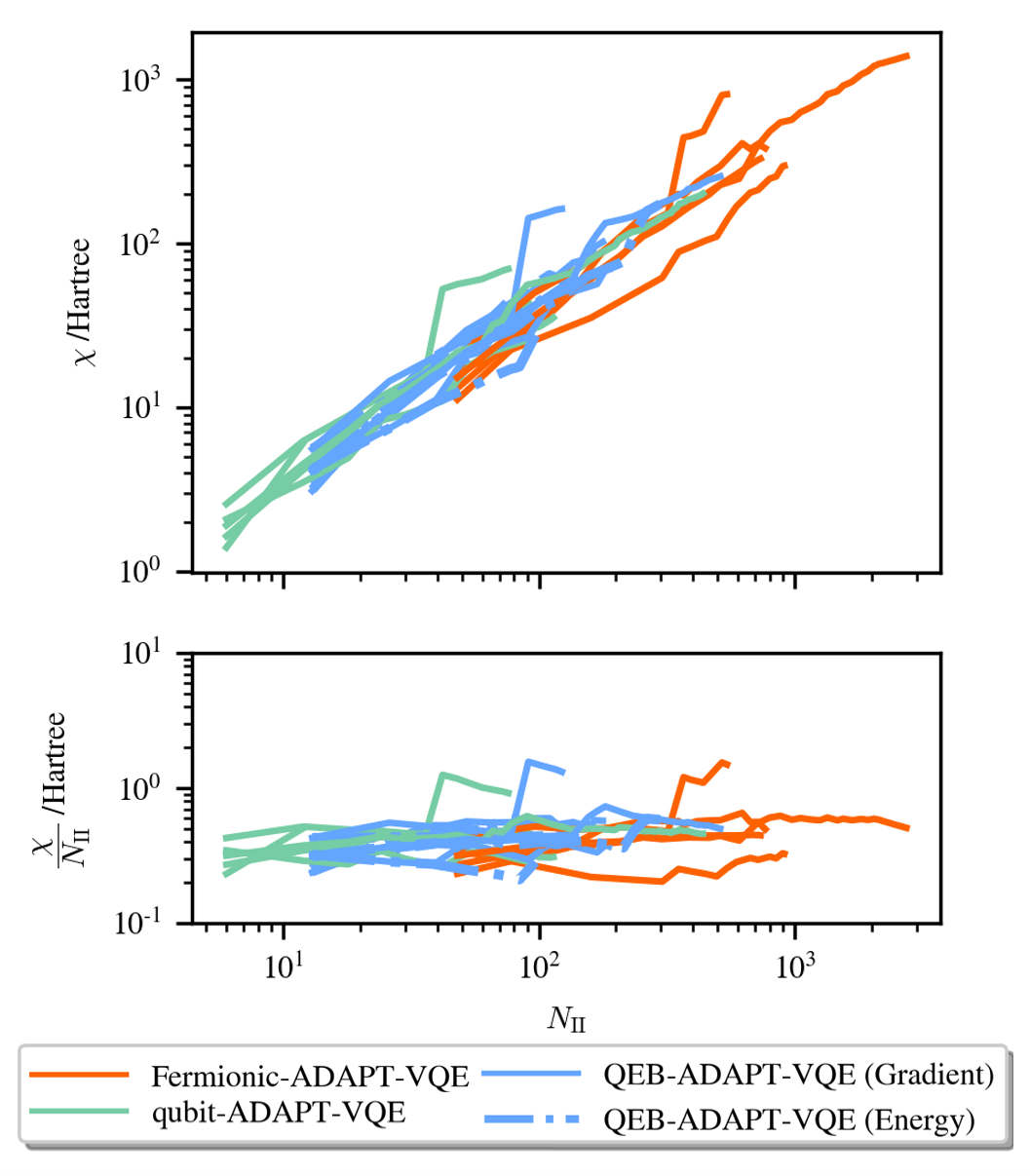}
\caption{
 Noise susceptibility (top) and average energy fluctuation  (bottom)  as functions of the number $\ncx$ of 
 CNOT gates for all molecules and algorithms reported in Sec.~\ref{1DPlotsSection}, at all circuit depths.}
\label{fig:scaling}
\end{figure}

\textit{Simplified computations:---}The energy expectation values underlying $\chi$ can be simulated with unitary 
operations on a state vector.
Such simulations are significantly simpler to perform than  density-matrix 
simulations.
Thus, we can more easily estimate the energy accuracy for small values of $p$:
\begin{linenomath} \begin{align}
\label{eq:EnergyFromSusceptibility}
\Delta E(p) \approx \chi p.
\end{align} \end{linenomath}
To test our method we compare Eq. \eqref{eq:EnergyFromSusceptibility} (black 
dotted lines) with some curves in 
Fig~\ref{fig:1DPlots}.
Equation \eqref{eq:EnergyFromSusceptibility}  estimates the 
simulated data  remarkably well.
Next, we use our method to estimate $\pc$ for molecules too large to study with 
our density-matrix simulator.  The estimates of $\pc$ for
$\mathrm{H}_2\mathrm{0}$ and $\mathrm{NH}_2^{-}$  are listed in the right-most 
section of Tab.~\ref{1DErrorTable}.
We stress that Eq.~\eqref{eq:EnergyFromSusceptibility} is an excellent predictor 
of 
$\Delta E(p)$ for the  gate-error probabilities $p\in[0,\pc]$ which allow for 
chemically-accurate simulations.

\textit{Scaling:---}The energy fluctuations are bounded by the spectral range of $H$: $\delta E \le 
\mathcal{E}_{\mathrm{max}} 
- \mathcal{E}_{0}$. Thus,   Eq.~\eqref{eq:NoiseSusceptibility} suggests that noise 
susceptibility grows linearly with  $\ncx$, as $\delta E$ is constant.
Fig.~\ref{fig:scaling} supports this claim. The curves indicate that $\chi 
\approxprop \ncx$ and $\delta E \approx \mathcal{O}(1)$, for a variety 
of molecules, ADAPT-VQE algorithms and circuit depths.
Combining these observations  with Eq.~\eqref{eq:EnergyFromSusceptibility}, we 
estimate that
\begin{linenomath} \begin{align}
\label{eq:scaling}
 \pc \approx\frac{\Delta E_C}{\delta E } \times  \frac{1}{\ncx} \approxprop \frac{1}{\ncx}.
\end{align} \end{linenomath}
where $\Delta E_C = 1.6\times 10^{-3}$ Hartree (chemical accuracy). This result is supported by 
recent results in condensed matter systems \cite{KagomeVQE}. The inverse proportionality between 
$\pc$ and $\ncx$ suggests that gate-error probabilities will have to reach extremely small values 
for useful chemistry calculations with VQE algorithms to be viable. Alternatively, we require \Change{improved} 
VQE algorithms with shallower circuits and fewer noisy (two-qubit) gates.

\subsection{Quantum Error Mitigation}

\label{sec:qem}
In the absence of error-corrected hardware, several strategies 
to mitigate the effect of noise have been suggested \cite{QEMReview, TemmeMitigation, Benjamin_2017,Benjamin_2021}.
Quantum error mitigation is a family of strategies which generally rely on knowledge of a circuit, noise model, 
or both to generate a set of modified circuits. Sampling from these circuits can generate a better estimate 
of the noiseless circuit's output \cite{QEMReview}. 
While these strategies have been demonstrated in simple VQE implementations \cite{HFOnSCQC,QEMVQESQC,QEMVQESQC2}, 
they suffer, in general, from exponential scaling of 
sample requirements with qubit number \cite{VQELimitations,EMScaling,EMScaling2}, 
potentially preventing their viability in useful NISQ VQE implementations. 
Indeed, leading reviews on quantum computational chemistry \cite{QCCReview}, 
state that `it seems unlikely that error-mitigation methods alone would enable 
more than a small multiplicative increase in the circuit depth.'
This unfavourable scaling has also been observed experimentally, where it has prevented the use of all but the most simple mitigation strategies \cite{IBMNature}.

\begin{figure*}[t]
  \centering
  \includegraphics[width=0.95\textwidth,trim=4 2 4 2,clip]{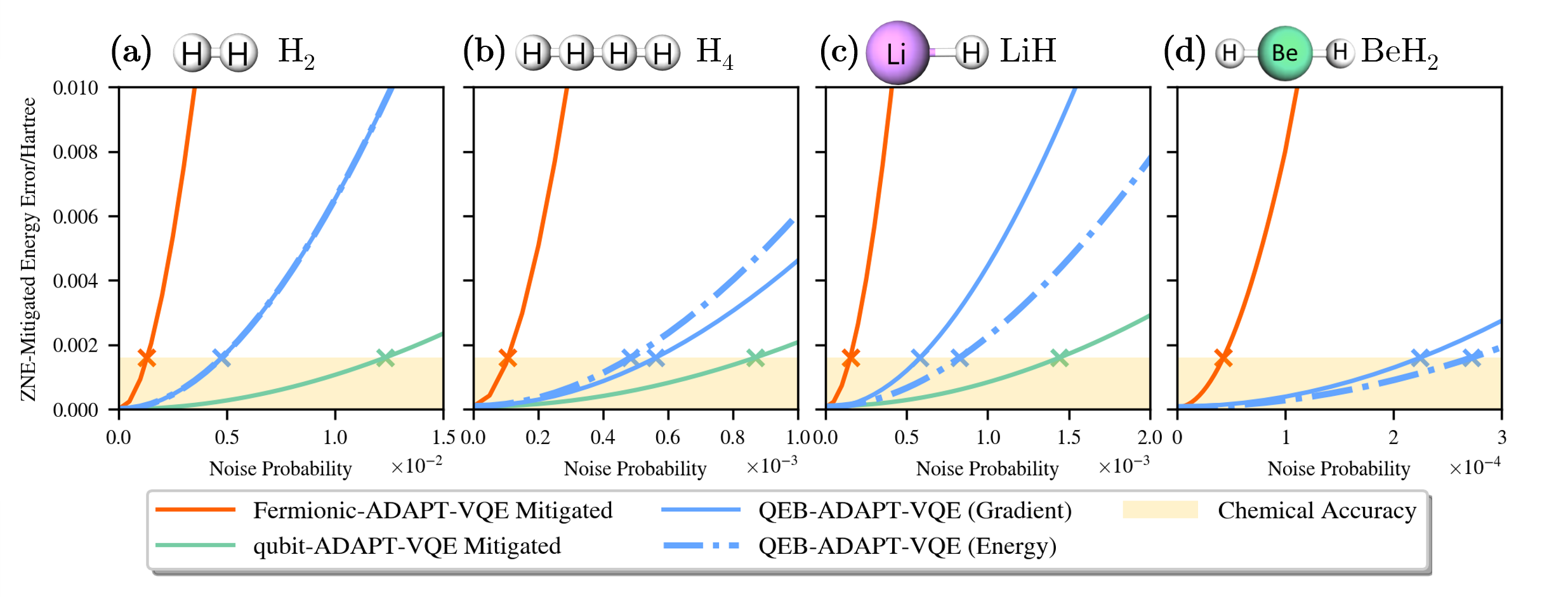}
  \caption{Plot representing the energy error (from FCI) using different 
  noise probabilities and VQE algorithms, for \Kieran{$\mathrm{H}_2$ (1 \AA) , $\mathrm{H}_4$ (1 \AA), $\mathrm{LiH}$ (1.546 \AA) and $\mathrm{BeH}_2$ (1.316 \AA),} 
  using linear zero-noise extrapolation \cite{TemmeMitigation,Benjamin_2017}. Energy accuracies 
  lower than chemical accuracy are highlighted by the yellow region. The crosses represent the intercepts 
  of the curves with this region.\label{fig:QEM_Combined}}
 \end{figure*}

The main goal of this work is to assess the required error rates for useful VQE implementations of 
molecules with more than 100 spin-orbitals. Due to the uncertainty around their scalability, as well as 
the unclear performance in the presence of time-dependent noise (particularly two-level-system defects \cite{TLSDefects} 
which drift in frequency), a study of this type should not include quantum error mitigation in its current form. Despite this, 
we believe it is relevant to extend our study to ascertain the \Change{maximally allowed gate-error probability $\pc$ to calculate 
molecular energies within chemical accuracy for an error-mitigation protocol with polynomial sampling 
overhead.}
To partially address this question, we repeat our numerical simulations using 
linear zero-noise extrapolation \cite{TemmeMitigation,Benjamin_2017} with a 
noise multiplication factor of 3.
Despite being biased and heuristic \cite{FutureQC}, we choose linear zero-noise extrapolation for its modest sampling 
overhead and numerical stability, which proved useful in recent 
large-scale demonstrations of error mitigation \cite{IBMNature}.

The results for $\mathrm{H}_2$, $\mathrm{H}_4$ and $\mathrm{LiH}$ are depicted in Fig.~\ref{fig:QEM_Combined}.
Compared to their counterparts in Fig.~\ref{fig:3DPlots}(a), we note the following:\
(i) The maximally allowed error probability increases by one or two orders of magnitude. This 
demonstrates the utility of error mitigation to make VQE more viable, especially 
for smaller molecules. 
(ii) The resulting energy error displays a roughly parabolic behaviour. This is 
expected from the series expansion of the depolarizing noise and indicates that 
further improvement (at an increased sampling overhead) may be possible by 
using higher-order extrapolations. 
(iii) We note that while all VQE algorithms display an increased noise 
resilience from error mitigation, their relative $\pc$-ranking does not change. 
This suggests that a VQE algorithm with higher noise 
resilience in the absence of error mitigation would remain more noise 
resilient when error mitigation is applied. 
Finally, we put the improved gate-error probabilities $\pc$ into 
context by plotting them as crosses in Fig.~\ref{fig:Table1}. 
Given the sharp decrease of $\pc$ with the problem size $N$, in the presence and absence of error mitigation, it is unlikely that error mitigation will improve $\pc$ sufficiently for useful system sizes, $N>100$. 
Ultimately, it remains an open question whether the unfavourable scaling of error mitigation prevents its use in realistic quantum-chemistry applications.

\section{Discussion}
\begin{figure}[t]
  \centering
  \includegraphics[width=0.5\columnwidth,trim=2 0 2 0,clip]{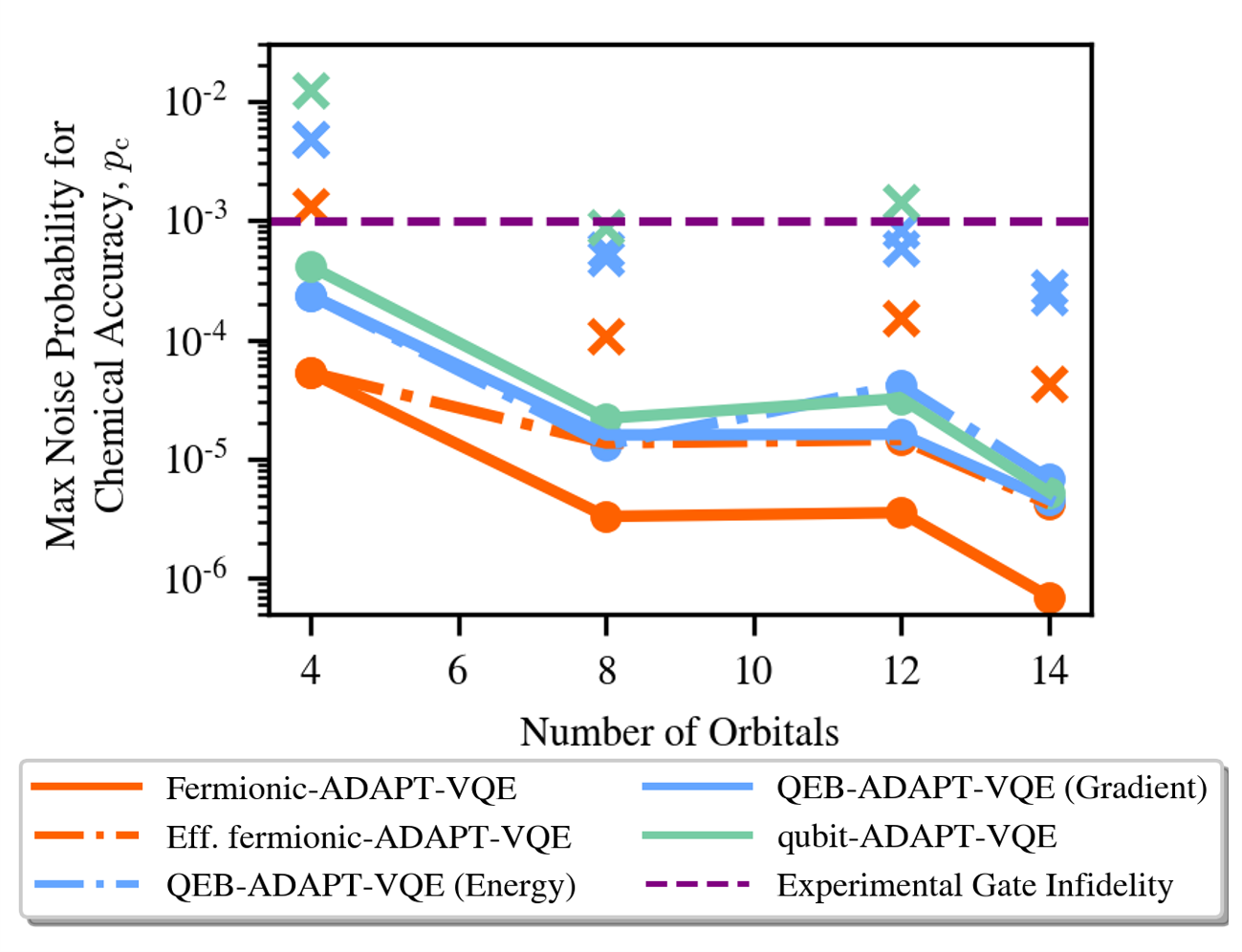}
  \caption{Plot representing the noise probability required to reach chemical accuracy, $p_c$, for different ansätze 
  and molecule sizes (number of orbitals). For molecules with the same number of orbitals, the mean probability is taken. The crosses and circles represent the noise probabilities required to reach 
  chemical accuracy with and without error mitigation, respectively. The data without error mitigation is taken from Table \ref{1DErrorTable}, and the data with error mitigation is taken from the crosses in Fig.~5. Additionally, a recent state-of-the-art two-qubit gate error rate 
  with superconducting qubits is shown in purple \cite{MITLL_Fluxonium}.
  \label{fig:Table1}}
\end{figure}

Any quantum algorithm aimed at near-term NISQ 
devices must be designed to tolerate some level of noise.
In this work, we \Kieran{numerically} quantify the maximally allowed depolarizing gate-error 
probabilities, $p_c$, required by leading gate-based VQEs to achieve chemically 
accurate energy estimates.
Based on numerical simulations, we reach five conclusions.
First, even the best-performing VQE algorithms require 
gate-error probabilities between $10^{-6}$ and $10^{-4}$, for the small molecules we assess. 
Such errors are 
at least an order of magnitude below state-of-the-art experiments 
\cite{SQCReview,IBMQuantumSystems} and the surface-code threshold 
\cite{SurfaceCodeReview,SurfaceCodeReview2}. If error mitigation is viable, the $p_{\mathrm{c}}$ values can be improved to $10^{-4}$ to $10^{-2}$ with linear zero-noise extrapolation.
Second, larger molecules tend to require longer ansatz-circuits and 
thus, lower gate-error probabilities, see Fig.~\ref{fig:Table1}. This is the case both with and 
without error mitigation.
Third, in the presence of noise, ADAPT-VQEs can tolerate 
approximately an order of magnitude greater gate-errors $p_c$ than equivalent 
fixed-ansatz VQEs, including those with the shortest ansatz circuits \cite{k-UpCCGSD,VQEReview}. 
Fourth, the more gate-efficient the ADAPT-VQE ansatz pool, the more 
noise resilient the algorithm. From a noise-resilience perspective, qubit 
excitations and Pauli-string excitations outperform fermionic excitations.
Fifth, the maximum gate-error probability allowed to reach chemical 
accuracy is roughly inversely proportional to the number of CNOT gates: $\pc 
\approxprop \ncx^{-1}$.
We now conclude this work with a couple of comments.

In this work, we quantify the maximally 
allowed gate-error probability of ADAPT-VQEs, UCCSD VQE and the leading fixed-ansatz VQE, k-UpCCGSD \cite{k-UpCCGSD}. The latter 
is chosen as due to its favourable circuit depth scaling with molecule size \cite{VQEReview}. 
This ignores plenty of other VQE algorithms which would benefit from similar 
studies in the future, as discussed in the main text \cite{iQCC,ClusterVQE,Ctrl-VQE,Disco-VQE, Overlap, PauliGrouping1, PauliGrouping2, PauliGrouping3,DoubleFactorisation1,DoubleFactorisation2}.
%
%

As opposed to a fault-tolerance threshold in error 
correction, the maximally allowed gate-error probability $\pc$ crucially depends on the size of the input problem, see Fig.~\ref{fig:Table1}. 
More specifically, $\pc$ tends to shrink as the number of spin orbitals $N$
increases. 
A key question for future research is to elucidate how fast $\pc$ decreases with  
$N$. 
Our numerical data in Fig.~\ref{fig:Table1} suggests an exponential scaling, both with and without error mitigation. 
Meanwhile, assuming VQEs achieve molecular ground-state energies with 
polynomially shallow circuits ($\ncx=\mathrm{poly}(N)$), Eq.~\eqref{eq:scaling} suggests a polynomial 
scaling. 
Having analytical expressions of the decrease of $\pc$ with the number of spin orbitals $N$, would inform us whether quantum advantage is at all feasible for input problems beyond 100 spin orbitals.


%

While this study is entirely focused on gate-errors, other sources of noise 
may also be relevant.
These include errors from state preparation and measurement as well as 
statistical noise due to sampling of expectation values from a limited number of 
shots.
As mentioned when justifying the noise model, errors due to state preparation 
and measurement tend to be smaller than the accumulated gate errors, and there 
are widely-implemented methods to compensate for them \cite{SPAM1, SPAM2, SPAM3, SPAM4, McWeeny, McWeenyBenchmark}.
However, in principle, measurement errors may lead to sub-optimal parameter values or 
operator choices during ansatz growth of ADAPT-VQE, \Kieran{which may prevent the algorithm from reaching the global minimum energy.}
A detailed analysis of such effects is left for future work.

While it is possible to sample any expectation value 
with $\epsilon$ accuracy in polynomially few shots \cite{QCCReview, VQEReview}, 
the scaling prefactor may lead to prohibitively large run-times \cite{HVA, VQERuntime1, 
VQERuntime2, VQEReview}. This issue is particularly acute for ADAPT-VQE algorithms, where each growth step 
requires shots for both parameter optimization and element selection. In this case, the number of necessary 
gradient measurements for each ansatz growth step is greater than that for VQE parameter optimization, 
by a factor which scales linearly in the number of qubits \cite{VirginiaGradients}. 
Holistic studies of VQE 
run-times \cite{HVA, VQERuntime1, VQERuntime2, VQEReview} provide predictions which vary greatly 
depending on the estimation methodology. The estimated run-times are often intractable without significant 
parallelization. The number of necessary measurements for parameter optimization can potentially be 
reduced via \Change{alternate} groupings of Pauli operators \cite{PauliGrouping1, PauliGrouping2, PauliGrouping3}, or tensor contraction 
of the Hamiltonian (such as by double factorization \cite{DoubleFactorisation1, DoubleFactorisation2, DoubleFactorisation3}). 
Despite this progress, run-time scaling remains a significant obstacle to overcome before ADAPT-VQEs can 
perform useful computations on real hardware. 
A balance must be found between run-time and the acceptable level of statistical noise. This is complicated by the combination of gate errors, measurement errors and statistical errors, which may affect VQEs adversely in a non-trivial way. 
We leave this as an open problem for the community as, in this work, we focus on the noise resilience of ADAPT-VQEs.

This work \Kieran{numerically} investigated the maximally 
allowed gate-error probability $\pc$ required to achieve chemically accurate 
predictions as a core metric of VQE performance. 
Similar to a fault-tolerance threshold in error correction, $\pc$ should provide 
a transparent metric to compare the noise resilience of VQEs as well as 
provide useful guidance for the experimental community. 
Having demonstrated that $\pc$ is between $10^{-4}$ and $10^{-6}$ for 
very small molecules (and worse for larger molecules), we 
conclude that quantum advantage in VQE-based quantum chemistry requires: 
(i) Substantially improved error mitigation, (ii) error correction, and/or (iii) 
significantly improved hardware in which gate errors are reduced by orders of 
magnitude. 

\textbf{Data Availability}
Data generated during the study is available upon request (E-mail:
kd437@cantab.ac.uk or ckl45@cam.ac.uk).

\textbf{Code Availability}
The code used for the simulations and analysis carried out for this work is available upon request from the authors (E-mail:
kd437@cantab.ac.uk or ckl45@cam.ac.uk).

\textbf{Acknowledgements}
We thank Hugo V. Lepage, Flavio Salvati, Frederico Martins, Joseph G. Smith, Wilfred Salmon, and the Hitachi QI team for 
useful discussions. NM and DRMAS acknowledge useful discussions with Tatsuya 
Tomaru, Saki Tanaka and Ryo Nagai.

\textbf{Author Contributions}
DRMAS and NM designed and supervised the project. YSY, CKL and KD wrote the simulation software. KD implemented the project 
and all numerical calculations in close collaboration with CKL.  CKL and NM developed the noise sensitivity analysis. 
All authors engaged in technical discussion and the writing of the manuscript.

\textbf{Competing Interests}
The authors declare no competing interests.

\bibliography{bibliography}{}

\clearpage

\end{document}


\begin{quantikz}
\newdimen\height
\setbox0=\hbox{$\mathcal D_{\vec p}$}
\height=\ht0 \advance\height by \dp0
\height=\dimexpr\height+0.6em\relax
\[
\begin{array}{c}
    \Qcircuit @C=1em @R=1em {
    \push{\hphantom{\text{Target qubit}}} &                              &     &     &                               &                           &          & \mbox{Noiseless Element}                               &          &                            &                                       &     &     &                            & \mbox{Noise Model}   &                            &     &     \\
    \push{\hphantom{\text{Target qubit}}} &                              &     &     &                               &                           &          &                                                        &          &                            &                                       &     &     &                            &                      &                            &     &     \\
    \push{\hphantom{\text{Target qubit}}} &                              &     &     &                               &                           &          & \mbox{Single Qubit Evolution $U\left(\theta\right)$}   &          &                            & \push{\hphantom{R_z(-\frac{\pi}{2})}} &     &     &                            &                      &                            &     &     \\
    \push{\hphantom{\text{Target qubit}}} &                              & \qw & \qw & \gate{R_z(\frac{\pi}{2})}     & \gate{R_x(\frac{\pi}{2})} & \ctrl{1} & \gate{R_x(\theta)}                                     & \ctrl{1} & \gate{R_x(-\frac{\pi}{2})} & \gate{R_z(-\frac{\pi}{2})}            & \qw & \qw & \qw                        & \qw                  & \qw                        & \qw & \qw \\
    \push{\hphantom{\text{Target qubit}}} & \lstick{\text{Target qubit}} & \qw & \qw & \push{\rule{0em}{\height}}\qw & \gate{R_x(\frac{\pi}{2})} & \targ    & \gate{R_z(\theta)}                                     & \targ    & \gate{R_x(-\frac{\pi}{2})} & \push{\rule{0em}{\height}}\qw         & \qw & \qw & \gate{\mathcal D_{\vec p}} & \qw                  & \gate{\mathcal D_{\vec p}} & \qw & \qw 
    {
        \gategroup{4}{4}{5}{11}{1em}{-}
        \gategroup{3}{4}{5}{11}{2.75em}{--}
        \gategroup{3}{14}{5}{16}{2.75em}{--}
    }}
\end{array}
\]
(a) Element by element noise model.
\[
\begin{array}{c}
    \Qcircuit @C=1em @R=1em {
    \push{\hphantom{\text{Target qubit}}} &                              &     &                               &                           &     &          &     & \mbox{Noise}                  &     &                               &                                     &          &     & \mbox{Noise}                  &     &                            &                        &                                &     \\
    \push{\hphantom{\text{Target qubit}}} &                              &     &                               & \mbox{Noiseless Gates}    &     &          &     & \mbox{Model}                  &     &                               & \mbox{\hphantom{m}Noiseless Gates}  &          &     & \mbox{Model}                  &     &                            & \mbox{Noiseless Gates} &                                &     \\
    \push{\hphantom{\text{Target qubit}}} &                              & \qw & \gate{R_z(\frac{\pi}{2})}     & \gate{R_x(\frac{\pi}{2})} & \qw & \ctrl{1} & \qw & \push{\rule{0em}{\height}}\qw & \qw & \push{\rule{0em}{\height}}\qw & \gate{R_x(\theta)}                  & \ctrl{1} & \qw & \push{\rule{0em}{\height}}\qw & \qw & \gate{R_x(-\frac{\pi}{2})} & \qw                    & \gate{R_z(-\frac{\pi}{2})}     & \qw \\
    \push{\hphantom{\text{Target qubit}}} & \lstick{\text{Target qubit}} & \qw & \push{\rule{0em}{\height}}\qw & \gate{R_x(\frac{\pi}{2})} & \qw & \targ    & \qw & \gate{\mathcal D_{\vec p}}    & \qw & \push{\rule{0em}{\height}}\qw & \gate{R_z(\theta)}                  & \targ    & \qw & \gate{\mathcal D_{\vec p}}    & \qw & \gate{R_x(-\frac{\pi}{2})} & \qw                    & \push{\rule{0em}{\height}}\qw  & \qw 
    {
        \gategroup{3}{4}{4}{7}{0.75em}{--}
        \gategroup{3}{9}{4}{9}{0.75em}{--}
        \gategroup{3}{11}{4}{13}{0.75em}{--}
        \gategroup{3}{15}{4}{15}{0.75em}{--}
        \gategroup{3}{17}{4}{19}{0.75em}{--}
    }}
\end{array}
\]
(b) Gate by gate noise model.
\end{quantikz}


\title{Supplementary Information for: Quantifying the effect of gate errors on variational quantum eigensolvers for quantum chemistry}
\author{Kieran Dalton}
\affiliation{Hitachi Cambridge Laboratory, J. J. Thomson Ave., Cambridge, CB3 
0HE, United Kingdom}
\affiliation{Cavendish Laboratory, Department of Physics, University of 
Cambridge, Cambridge, CB3 0HE, United Kingdom}
\affiliation{Current Address: Department of Physics, ETH Z\"urich, CH-8093 Z\"urich, 
Switzerland}
\author{Christopher K. Long}
\affiliation{Hitachi Cambridge Laboratory, J. J. Thomson Ave., Cambridge, CB3 
0HE, United Kingdom}
\affiliation{Cavendish Laboratory, Department of Physics, University of 
Cambridge, Cambridge, CB3 0HE, United Kingdom}

\author{Yordan S. Yordanov}
\affiliation{Hitachi Cambridge Laboratory, J. J. Thomson Ave., Cambridge, CB3 
0HE, United Kingdom}
\affiliation{Cavendish Laboratory, Department of Physics, University of 
Cambridge, Cambridge, CB3 0HE, United Kingdom}

\author{Charles G. Smith}
\affiliation{Hitachi Cambridge Laboratory, J. J. Thomson Ave., Cambridge, CB3 
0HE, United Kingdom}
\affiliation{Cavendish Laboratory, Department of Physics, University of 
Cambridge, Cambridge, CB3 0HE, United Kingdom}

\author{Crispin H. W. Barnes}
\affiliation{Cavendish Laboratory, Department of Physics, University of 
Cambridge, Cambridge, CB3 0HE, United Kingdom}

\author{Normann Mertig}
\affiliation{Hitachi Cambridge Laboratory, J. J. Thomson Ave., Cambridge, CB3 
0HE, United Kingdom}
\author{David R. M. Arvidsson-Shukur}
\affiliation{Hitachi Cambridge Laboratory, J. J. Thomson Ave., Cambridge, CB3 
0HE, United Kingdom}

\maketitle
\onecolumngrid
\section*{Supplementary Note 1: Pseudocode for Iterative Growth Algorithms}

Below we present pseudocode for the two main categories of iterative VQE 
algorithms we present in this paper: those using gradient-based decision 
rules and those using energy-based decision rules. These algorithms are 
summarized in the flowchart, Fig.~1, and covered in more 
detail in Sec.~II A.

\begin{figure}[h]
  \begin{algorithm}[H]
      \caption{QEB-ADAPT (Energy-based decision 
rule)}\label{QEBADAPTVQEPseudocode}
      \input{Figures/QEBpseudocode.tex}
\end{algorithm}
\end{figure}
\clearpage
\begin{figure}[h]
    \begin{algorithm}[H]
        \caption{Fermionic-ADAPT/qubit-ADAPT/QEB-ADAPT (Gradient-based 
decision rule)}\label{ADAPTVQEPseudocode}
        \input{Figures/ADAPTpseudocode.tex}
\end{algorithm}
\end{figure}

\section*{Supplementary Note 2: Noisy Optimization and Growth}

To justify the noiseless growth and parameter optimization of our 
ansätze, we must demonstrate that the noiselessly-optimized parameters, and the 
resulting energies, differ little from the parameters and energies 
optimized with noise. This claim is supported by the results in Refs. 
\cite{QAOAParameters,QAOAParametersFollowup,NoiseImpact, BiamonteCoherentError}, which describe how 
increasing noise dampens the energy landscape without moving the 
location of the global minimum. In Supplementary Fig.~\ref{fig:NoisyOptimisationTest}, we plot the 
energies obtained using noiselessly-grown ansätze for $\mathrm{H}_4$ 
(1\AA\ interatomic separation), $\mathrm{LiH}$ and $\mathrm{HF}$. The solid lines represent 
energies obtained using noiselessly-optimized parameters, while the dotted lines 
represent energies obtained using parameters optimized in the presence of noise. 
Parameters are optimized using the gradient-free Nelder-Mead optimizer \cite{Nelder-Mead}.
These noisily-optimized lines lie beneath the noiselessly-optimized lines over the 
noise probability range we assess (that around which chemical accuracy is achieved).
Thus, there is not a significant shift of the global minimum in parameter space, 
and hence our approach is justified.

\begin{figure*}[h]
\includegraphics[width=\textwidth, trim=5 2 7 2, clip]{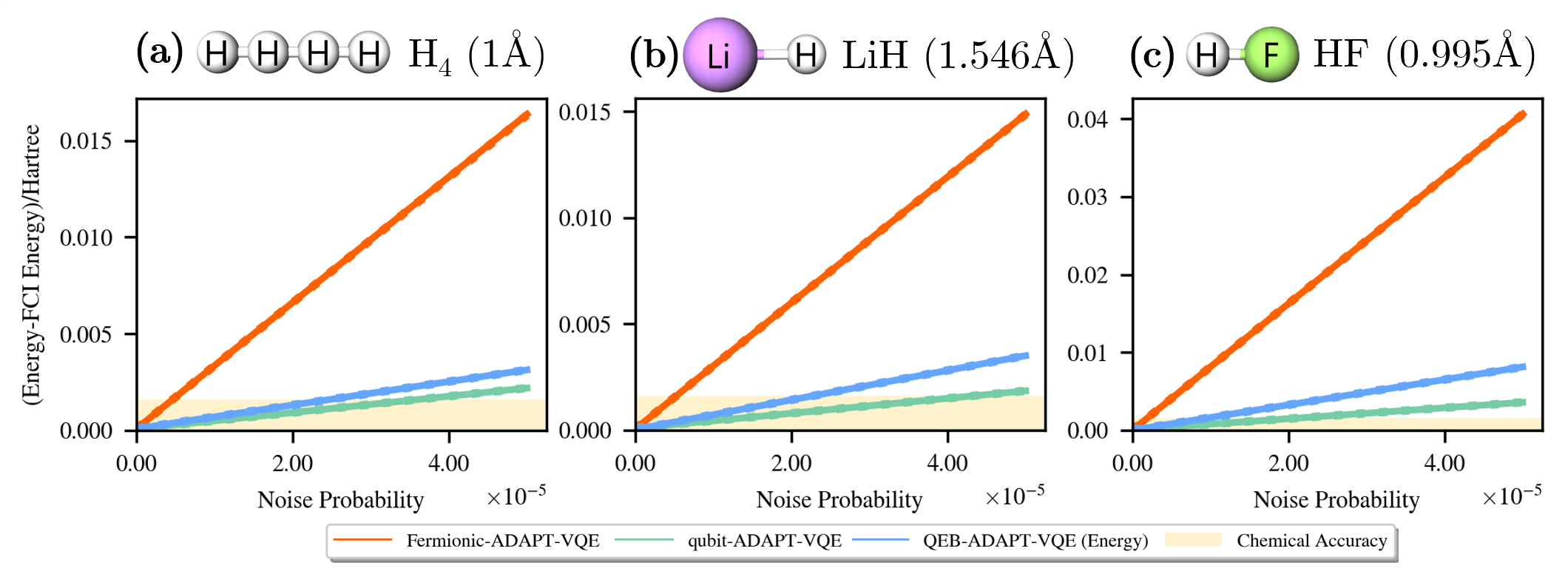}
\caption{Energies calculated using (solid lines) noiselessly-optimized and (dotted lines) noisily-optimized 
parameters, using the same ansätze. \label{fig:NoisyOptimisationTest}}
\end{figure*}

Now, we must consider ansätze grown in the presence of noise. Intuitively, there are two possible effects of 
this noisy growth. First, the damping of the energy landscape will result in any algorithm halting criterion to 
be satisfied sooner, resulting in shorter ansätze. This is preferential in the presence of noise, and we have 
investigated early ansatz truncation thoroughly in Sec.~II D. Second, noise may result in the selection of an 
element which is not noiselessly optimal. This is likely to favour shorter elements (single excitations, rather 
than double excitations), and may harm convergence. To investigate this, we executed the QEB-ADAPT-VQE algorithm 
in the presence of noise for the $\mathrm{H_4}$ molecule, for a noise probability values of $4 \times 10^{-6}$, $4 \times 10^{-5}$ and $4 \times 10^{-4}$.
For a larger value of $4 \times 10^{-3}$, the noisy growth failed.

We note that, in the presence of noise, the analytic formula used to calculate gradients breaks down. Thus, 
gradients were calculated using the finite differences method. Here, when calculating the gradient of a new 
ansatz element, we first append the element and calculate the noisy energy expectation value for a parameter 
value of zero. Then, we repeat this for a small finite parameter value, and use the difference to estimate the 
gradient. The element is first appended with a parameter value of zero to emulate the noise induced by the element 
and prevent the discrete jump in energy values which occurs upon the addition of any element to the ansatz 
(due to the additional noise burden). Parameters are then optimized using the gradient-based BFGS optimizer \cite{BFGS}.

\begin{figure*}[h]
  \includegraphics[width=\textwidth, trim=2 0 2 0, clip]{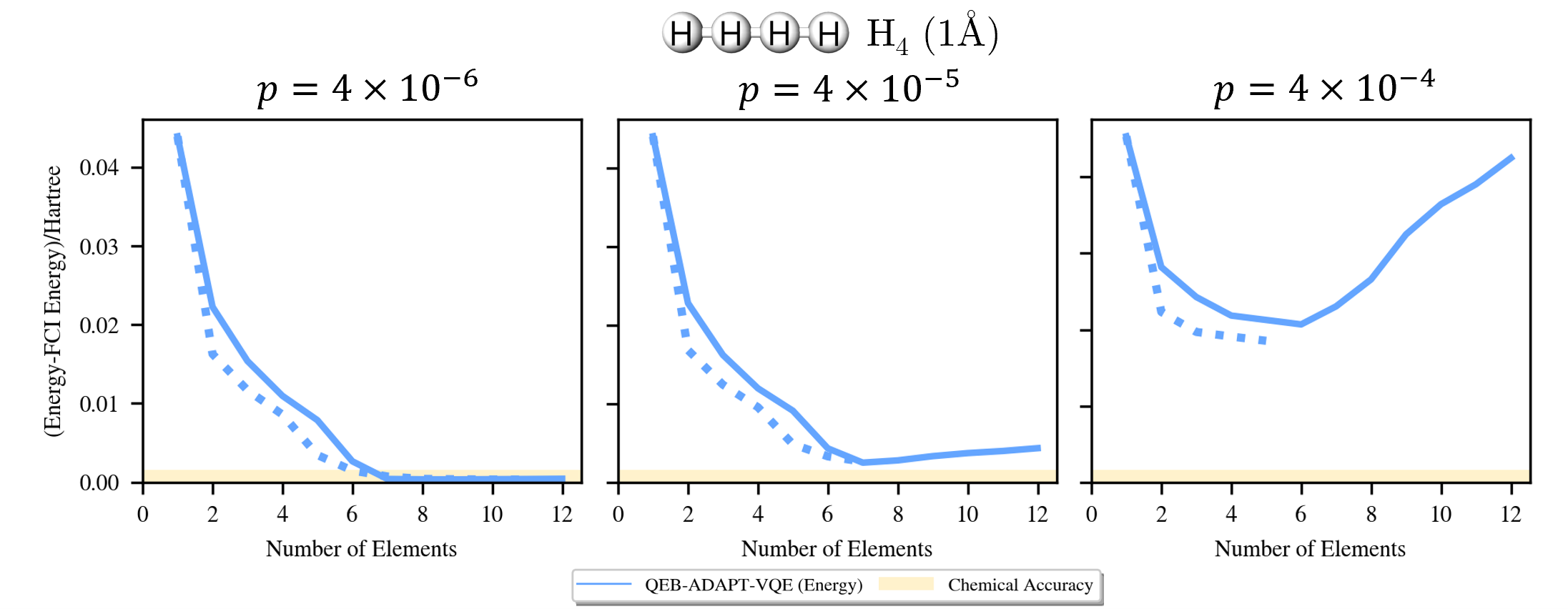}
  \caption{Energies calculated using (solid lines) noiselessly-grown and (dotted lines) noisily-grown 
  ansätze, with parameters optimized using gradients estimated with the finite difference method. \label{fig:NoisyGrowthTest}}
  \end{figure*}

Our results for noisy growth and optimization are shown in Fig.~\ref{fig:NoisyGrowthTest}. Growth and optimization in the presence of noise results in a 
different ansatz circuit which, in this case, is more noise-resilient. However, the improvement is not sufficient to make the algorithm feasible on NISQ devices. 
Additionally, as expected, the ansatz grown with noise is shorter than that grown without noise, and the final length is greater for lower noise probabilities.

\clearpage

\section*{Supplementary Note 3: Comparison of Noise Models}

For molecules larger than $\mathrm{H}_2$, we improve simulation 
performance by applying noise after each ansatz element, rather than after each 
CNOT gate. The post-element noise is applied to the qubits targeted by CNOT 
gates in the preceding ansatz element, with the number of applications equal to 
the number of times that qubit was targeted in the element. This is illustrated 
in Supplementary Fig.~\ref{fig:noise model} for a single qubit excitation. The performance improvement from applying noise after each ansatz element, rather than after each 
gate, is significant. However, this modification results in an underestimation of the energy error, highlighted in Supplementary Fig.~\ref{fig:EBEvsGBG}. This underestimation does not alter the performance hierarchy of the fermionic-ADAPT-VQE, qubit-ADAPT-VQE and QEB-ADAPT-VQE.

\begin{figure}[h]
  \includegraphics[width=\textwidth, trim=5 0 0 0, clip]{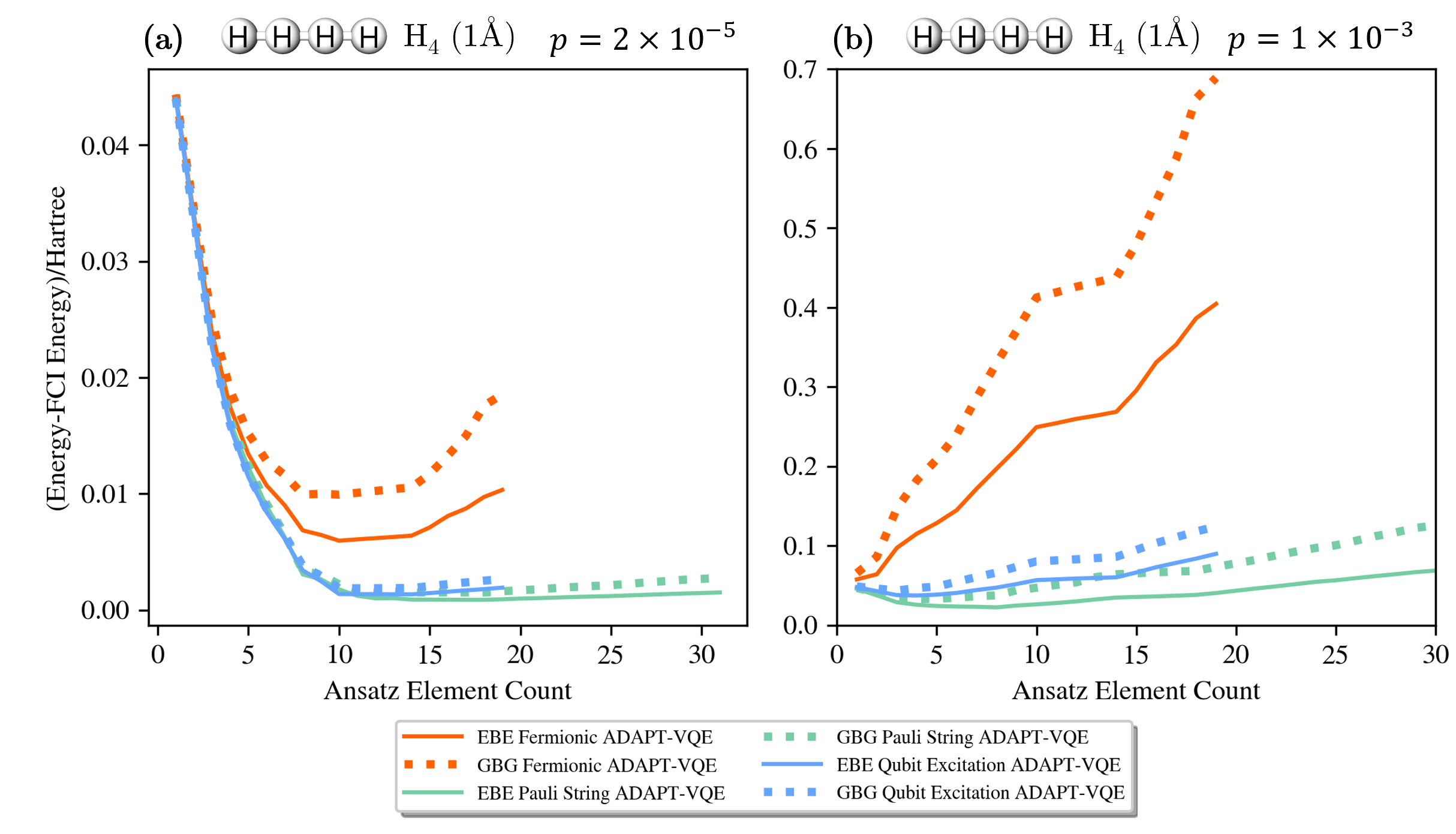}
  \caption{Energy error for $\mathrm{H}_4$, simulated using the fermionic-ADAPT-VQE, qubit-ADAPT-VQE and QEB-ADAPT-VQE (with the gradient decision rule). The solid lines are energies evaluated using noise applied after each ansatz element (the element-by-element, or EBE, method), while the dotted lines are energies evaluated using noise applied after each CNOT gate (the gate-by-gate, or GBG, method). The right figure uses a realistic noise probability of 0.1\%, whereas the left figure uses a noise probability representative of the values used in our simulations. \label{fig:EBEvsGBG}}
\end{figure}
\clearpage

\begin{figure*}[h]
\includegraphics[width=\textwidth, trim=0 0 0 5, clip]{Figures/Final/Supp_Fig4.pdf}
\caption{Noisy circuit diagrams of single qubit evolutions for each of the noise models used: (a) element by element noise model; (b) gate by gate noise model.}\label{fig:noise model}
\end{figure*}

\section*{Supplementary Note 4: Additional VQE Ansätze}

\begin{figure*}[h!]
  \hspace*{-1cm}
  \includegraphics[width=\textwidth,trim=2 0 2 0,clip]{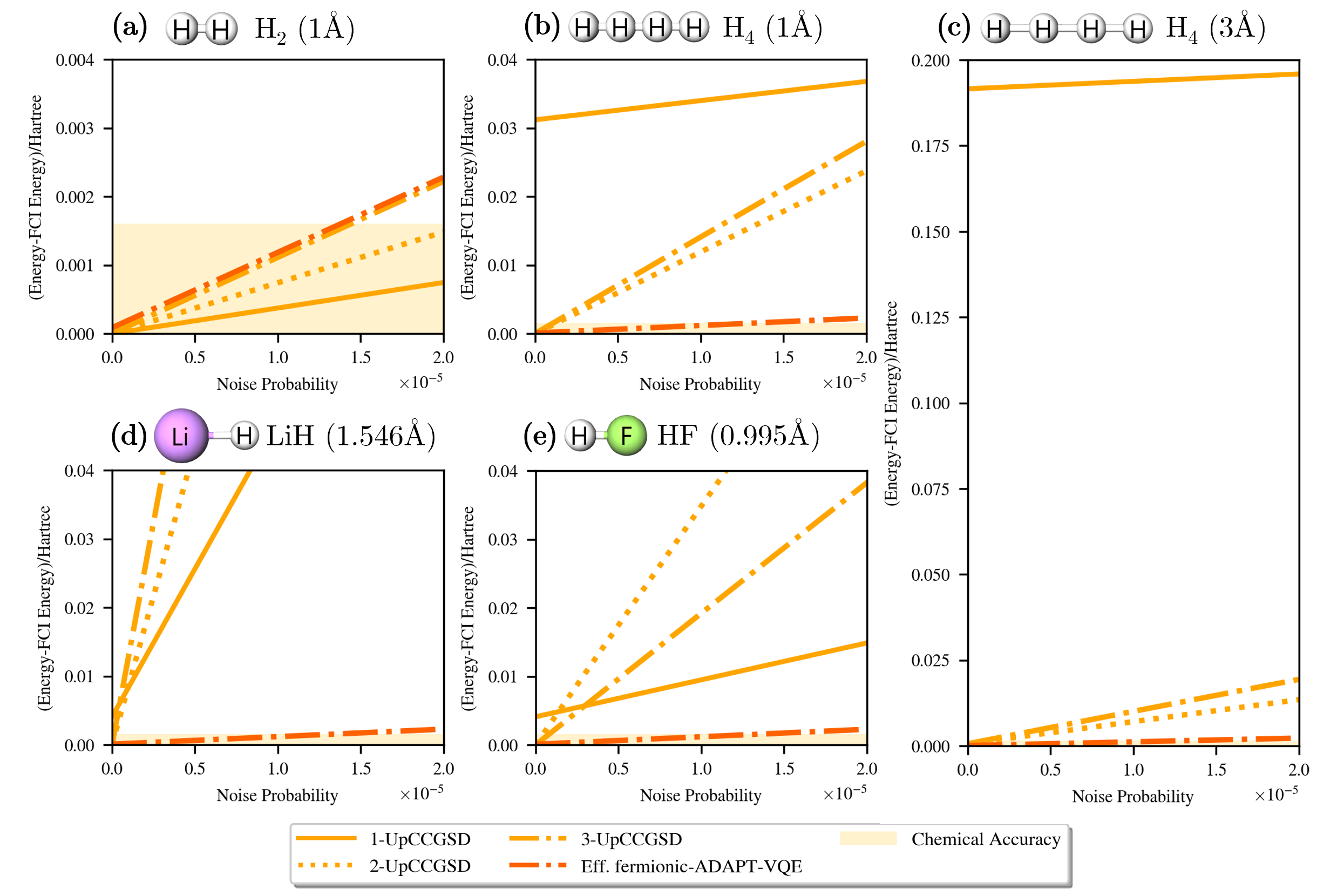}
  \caption{Energy accuracy as a function of gate-error probability for $\mathrm{H}_{2}$, 
  $\mathrm{H}_4$ (at 1\AA\ and 3\AA\ interatomic separation), $\mathrm{LiH}$ and $\mathrm{HF}$. 
  Energy accuracies lower than chemical accuracy are highlighted by the yellow region. The orange lines are  
  the energies calculated using the k-UpCCGSD ansatz, for $k=1$, $k=2$ and $k=3$ \cite{k-UpCCGSD}. The red line is the 
  fermionic-ADAPT-VQE ansatz, discussed in the main text.  All ansätze shown use efficient fermionic elements.\label{fig:kUpCCGSD}}
\end{figure*}

In the main text, we simulated the performance of both fixed (UCCSD) and iteratively-grown 
(ADAPT-VQE) VQE ansätze. Here, we extend our range of assessed ansätze to further fixed variants. In Fig.~\ref{fig:kUpCCGSD}, 
we present the results for the k-UpCCGSD ansatz  \cite{k-UpCCGSD}, with $k=1$, $k=2$ and $k=3$ repetitions, using efficient representations 
of fermionic excitation evolutions \cite{YordanEfficientQuantumCircuits}. Increasing the number of repetitions increases circuit depth while 
improving performance for strongly-correlated systems, 
and the k-UpCCGSD ansatz is recommended by a recent VQE review due to the linear scaling of circuit depth with qubit number \cite{VQEReview}. 
Despite this, we observe that the algorithm does not 
reach chemical accuracy for $k=1$, and displays poor noise resilience for $k=2$ and $k=3$ repetitions. The energy error is significantly greater 
than that of fermionic-ADAPT-VQE, 
justifying our focus on ADAPT-VQE methods in the main text.

\clearpage

\section*{Supplementary Note 5: Additional Colour Plots}
Here, we present excess colour plots for the algorithms not included in Fig.~3.

\begin{figure*}[h!]
  \hspace*{-1cm}
  \includegraphics[width=1\textwidth,trim=5 0 7 0,clip]{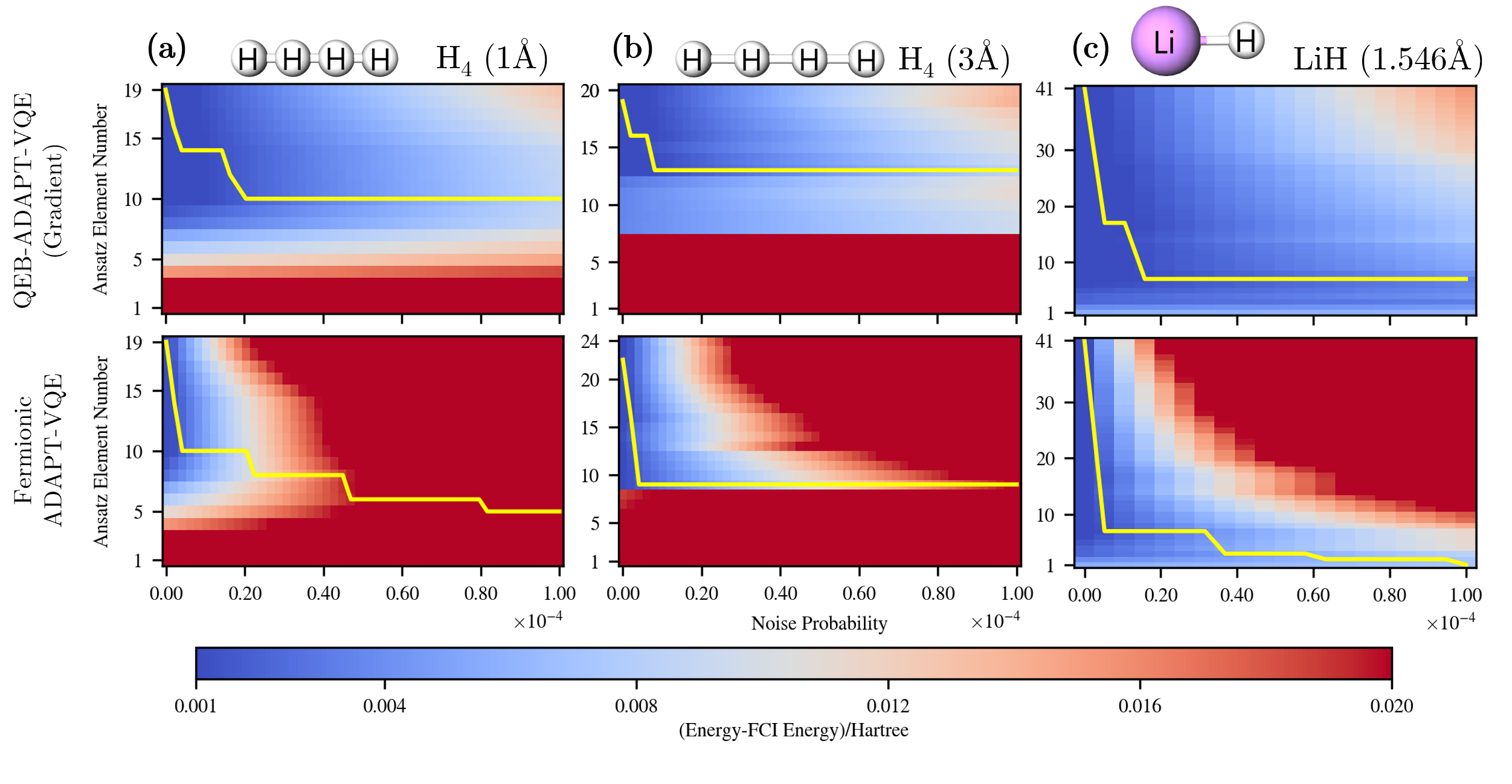}
  \caption{Colour plots representing the energy error (from FCI) at different 
noise probabilities and ansatz lengths, for three different molecules. Ansätze 
are grown using the two different iterative-growth methods excluded from Fig.~3: fermionic-ADAPT-VQE (with inefficient elements) and QEB-ADAPT-VQE (with the gradient-based decision rule). The yellow line on the colour plot highlights the ansatz length which minimizes energy error for each noise 
probability.\label{fig:3DPlotsExtra}}
\end{figure*}

\pagebreak
\section*{Supplementary Note 6: Noise susceptibility}
\label{app:NoiseSusceptibility}

\renewcommand{\gate}{\mathcal{G}}
\newcommand{\depolarize}{\mathcal{D}}
\newcommand{\Xgate}{\mathcal{X}}
\newcommand{\Ygate}{\mathcal{Y}}
\newcommand{\Zgate}{\mathcal{Z}}

\textit{Notation}:---To derive the expressions on noise susceptibility in Sec.~II E, we 
now introduce some notation.
Let $\circuit(p)$ denote the completely-positive trace-preserving (CPTP) map 
which describes the noisy dynamics of a VQE ansatz.
For brevity, we do not explicitly highlight the dependence on circuit parameters $\theta_1,\ldots,\theta_n$.
Without noise 
\begin{linenomath} \begin{align}
 \circuit(0) = \gate_R\circ\ldots\circ\gate_r\circ\ldots\circ\gate_1,
\end{align} \end{linenomath}
where the noiseless CPTP maps of the unitary gates unitary gates $G_1,\ldots,G_R$ are represented by $\gate_1,\ldots,\gate_R$.
$\circ$ denotes CPTP map composition. Following the noise model in Sec.~II B, we 
express the noisy circuit as
\begin{linenomath} \begin{align}
 \label{eq:AppNoisyCircuit}
 \circuit(p) = \Lambda_R(p)\circ\gate_R\circ 
               \ldots\circ
               \Lambda_r(p)\circ\gate_r\circ
               \ldots\circ
               \Lambda_1(p)\circ\gate_1,
\end{align} \end{linenomath}
where $\Lambda_r(p)$ denotes a noise channel.
Further, we define $\cnotset$ to denote the set of indices for which $G_r$ is 
a controlled-NOT gate and $i_r$ to be the controlled qubit in the $r$th gate 
operation.
This allows us to specify the noise operators $\Lambda_r$ for $r=1,\ldots,R$ as
\begin{linenomath} \begin{align}
    \Lambda_r(p) = \begin{cases}
                    \mathds{I}         & \text{ if } r\notin\cnotset\\
                    \depolarize(p,i_r) & \text{ if } r\in\cnotset.
                   \end{cases}
\end{align} \end{linenomath}
Here, $\depolarize(p,i)$ denotes a depolarizing channel acting on qubit $i$.
For later reference, we define the CPTP map induced by the Pauli operators $X_i, 
Y_i, Z_i$ acting on qubit $i$ as $\Xgate_i, \Ygate_i, \Zgate_i$. Thus,
\begin{linenomath} \begin{align}
 \depolarize(p,i) \equiv (1-p)\mathds{I} 
 + \frac{p}{3}\left(\Xgate_i+\Ygate_i+\Zgate_i\right).
\end{align} \end{linenomath}
Further, we denote the expectation value of the Hamiltonian $H$ given an 
input state $\rho_0$ and a circuit $\circuit(p)$ as
\begin{linenomath} \begin{align}
 \label{eq:AppEnergyExpectation}
 E(p) = \Trace\left[H\,\circuit(p)[\rho_0]\right].
\end{align} \end{linenomath}
Finally, using the full-configuration-interaction energy $E_{\mathrm{FCI}}$ as the true ground state $\mathcal{E}_0$ of $H$, we recall that energy accuracy is
\begin{linenomath} \begin{align}
 \Delta E(p) = E(p) - E_{\mathrm{FCI}}.
\end{align} \end{linenomath}

\textit{Noise susceptibility}:---
We define noise susceptibility $\chi[\circuit(p)]$ of  $\circuit(p)$ 
to be the linear response of the energy accuracy of the unitary circuit at 
$p=0$, defined by
\begin{linenomath} \begin{align}
 \chi[\circuit(p)]
  := \left.\frac{\partial \Delta E(p)}{\partial p}\right|_{p=0}
   = \left.\frac{\partial E(p)}{\partial p}\right|_{p=0}.
\end{align} \end{linenomath}
Inserting Eq.~\eqref{eq:AppEnergyExpectation}  allows us to write
\begin{linenomath} \begin{align}
\label{eq:AppNoiseSusceptibilityIntermediate}
 \chi[\circuit(p)] &= \left.\frac{\partial
                      \Trace\left[H\,\circuit(p)[\rho_0]\right]
                      }{\partial p}\right|_{p=0}
                   =                     
\Trace\left[H\,\left.\frac{\partial\circuit(p)}{\partial p}\right|_{p=0}
                      [\rho_0]\right],
\end{align} \end{linenomath}
where we used linearity of the trace and the CPTP map as well as the fact that 
neither $H$ nor $\rho_0$ depend on $p$.
Further, inserting Eq.~\eqref{eq:AppNoisyCircuit} and applying the product rule, we establish that 
\begin{linenomath} \begin{align}
 \left.\frac{\partial\circuit(p)}{\partial p}\right|_{p=0} & = 
 \sum_{r=1}^{R}
  \left.\Lambda_R(p)\right|_{p=0}\circ\gate_R\circ 
               \ldots\circ
               \left.\frac{\partial\Lambda_r(p)}{\partial p}\right|_{p=0}
               \circ\gate_r\circ
               \ldots\circ
               \left.\Lambda_1(p)\right|_{p=0}\circ\gate_1.
\end{align} \end{linenomath}
Since $\left.\Lambda_r(p)\right|_{p=0}=\mathds{I}$, $\forall r=1,\ldots,R$,
this implies
\begin{linenomath} \begin{align}
 \label{eq:Cderivative}
\left.\frac{\partial\circuit(p)}{\partial p}\right|_{p=0} & = 
 \sum_{r=1}^{R}
  \gate_R\circ 
               \ldots\circ
               \left.\frac{\partial\Lambda_r(p)}{\partial p}\right|_{p=0}
               \circ\gate_r\circ
               \ldots
               \circ\gate_1.
\end{align} \end{linenomath}          
Next, we compute the derivative of $\Lambda_r(p)$,
\begin{linenomath} \begin{align}
\left.\frac{\partial\Lambda_r(p)}{\partial p}\right|_{p=0}
    = \begin{cases}
     0     & \text{ if } r\notin\cnotset\\
     \frac{1}{3}\left(\Xgate_{i_r}+\Ygate_{i_r}+\Zgate_{i_r}\right) - 
\mathds{I} & \text{ if } r\in\cnotset,
       \end{cases}
\end{align} \end{linenomath}
and insert this into Eq.~\ref{eq:Cderivative}
\begin{linenomath} \begin{align}
 \label{eq:AppDiffNoisyCircuit}
 \left.\frac{\partial\circuit(p)}{\partial p}\right|_{p=0} = 
 \sum_{r=\cnotset}
  \gate_R\circ\ldots\circ\gate_{r+1}\circ
\left(\frac{1}{3}\left(\Xgate_{i_r}+\Ygate_{i_r}+\Zgate_{i_r}\right) - 
    \mathds{I}\right)
               \circ\gate_r\circ
               \ldots\circ\gate_1.
\end{align} \end{linenomath}
Finally, denoting the set of CPTP maps acting on qubit $i$ as 
\begin{linenomath} \begin{align}
 P(i) = \{\Xgate_i,\Ygate_i,\Zgate_i\},
\end{align} \end{linenomath}
substituting Eq.~\eqref{eq:AppDiffNoisyCircuit} into Eq.~\eqref{eq:AppNoiseSusceptibilityIntermediate} and using the linearity of the trace, we obtain the noise susceptibility as
\begin{linenomath} \begin{align}
 \chi[\circuit(p)]
  := \sum_{r\in\cnotset}\frac{1}{3}\sum_{\Sigma_{i_r}
\in P(i_r)}\Trace\left[H\,
    \gate_R\circ\ldots\circ\gate_{r+1}\circ
\left(\Sigma_{i_r}-\mathds{I}\right)
               \circ\gate_r\circ
               \ldots\circ\gate_1[\rho_0]\right].
\end{align} \end{linenomath}
To interpret this expression, we decompose it further.
Noting that $\Sigma_i$ is a CPTP map induced by the unitary Pauli operator 
$\sigma_i\in\{X_i,Y_i,Z_i\}$ acting on qubit $i$, we introduce the expectation 
value 
\begin{linenomath} \begin{align}
 E(\circuit, \sigma, r, i) :=  \Trace \left[H\,
    \gate_R\circ\ldots\circ \gate_{r+1} \circ
\Sigma_{i}\circ\gate_r\circ \ldots\circ\gate_1[\rho_0]\right].
\end{align} \end{linenomath}
$E(\circuit, \sigma, r, i)$ can  be re-expressed in terms of noiseless unitary operations:
\begin{linenomath} \begin{align}
 E(\circuit, \sigma, r, i) =  \Trace \left[H\,
    G_R\ldots G_{r+1} \sigma_{i} G_r \ldots G_1\rho_0
    G_1^{\dagger}\ldots G_{r}^{\dagger} \sigma_{i} G_{r+1}^{\dagger} \ldots 
G_R^{\dagger} \right].
\end{align} \end{linenomath}
This expectation value is obtained by perturbing the originally 
intended noiseless ansatz circuit with a Pauli operator $\sigma$ after gate $r$ 
on qubit $i$.
In a similar manner we introduce
\begin{linenomath} \begin{align}
 E(\circuit) & :=  \Trace \left[H\,\gate_R\circ\ldots\circ\gate_r\circ 
\ldots\circ\gate_1[\rho_0]\right], \\
& =  \Trace \left[H\,
    G_R\ldots G_r \ldots G_1\rho_0
    G_1^{\dagger}\ldots G_{r}^{\dagger} \ldots G_R^{\dagger} \right].
\end{align} \end{linenomath}
This is the energy expectation value of the ideal noiseless 
ansatz.
With these expressions in place, we re-express the noise susceptibility as
\begin{linenomath} \begin{align}
 \label{eq:AppNoiseSusceptibility}
 \chi[\circuit(p)]
  := \left|\cnotset\right| \left(
  \frac{1}{\left|\cnotset\right|}\sum_{r\in\cnotset}\frac{1}{3}\sum_{\sigma 
\in {X,Y,Z}}
  \left[E(\circuit, \sigma, r, i_r)-E(\circuit)\right] \right).
\end{align} \end{linenomath}

\textit{Discussion:---}(i) Energy fluctuations:\ We interpret noise susceptibility as follows.
$E(\circuit)$ is the expectation value of the ideal ansatz. It will typically be close to the true ground state $\mathcal{E}_O$. 
$E(\circuit, \sigma, r, i_r)$, on the other hand, could be a random perturbation 
towards anywhere in the energy spectrum. Thus, it is reasonable to assume 
that $E(\circuit, \sigma, r, i_r)$ would typically be larger than 
$E(\circuit)$. We interpret the difference of the two 
contributions
\begin{linenomath} \begin{align}
 \delta E(\circuit, \sigma, r, i_r) = E(\circuit, \sigma, r, 
i_r)-E(\circuit).
\end{align} \end{linenomath}
as an energy excitation induced by the noise on the controlled qubit $i$ of the 
controlled NOT gate executed in position $r$.
Taking this interpretation one step further, we consider the large bracket in 
Eq.~\eqref{eq:AppNoiseSusceptibility} as an average of all $X,Y,Z$ perturbations 
along the circuit $\circuit$ after each of the controlled NOT gates 
$r\in\cnotset$, defined as
\begin{linenomath} \begin{align}
 \delta E(\circuit) := 
  \frac{1}{\left|\cnotset\right|}\sum_{r\in\cnotset}\frac{1}{3}\sum_{\sigma 
\in {X,Y,Z}}
  \left[E(\circuit, \sigma, r, i_r)-E(\circuit)\right].
\end{align} \end{linenomath}

(ii) Computability:\ The fact that noise susceptibility can entirely be 
expressed in terms of unitary computations instead of CPTP maps facilitates an advantage for computation.
First, a full density matrix computation requires storing this full density matrix in memory.
For N qubits this requires storing a matrix of $(2^N)^{2}$ complex numbers.
This is usually a key bottleneck which limits tractable calculations.
Using the unitary computation requires propagating $|\cnotset|$ complex state 
vectors each of which requires storing only $2^N$ complex numbers at a time.
While still computationally expensive, this allows computing noise 
susceptibility for larger molecules, such as $\mathrm{H}_2\mathrm{O}$.
Finally, as seen in Fig.~2, for $p\lesssim\pc$ energy accuracy 
can be well approximated by the noise susceptibility as 
\begin{linenomath} \begin{align}
 \Delta E(p) \approx \chi p + \oforder(p^2).
\end{align} \end{linenomath}
This allows for approximating the threshold below which 
 chemical accuracy can be reached without error mitigation as
\begin{linenomath} \begin{align}
 p_c \approx \frac{1.6\times10^{-3}\mathrm{ Hartree}}{\chi_{\circuit}} = 
\frac{1.6\times10^{-3}\mathrm{ Hartree}}{\delta 
E(\circuit)}\frac{1}{|\cnotset|}.
\end{align} \end{linenomath}
This relation corroborates that $ p_c$
scales in inverse proportion with the number of two-qubit gates.
Ultimately, this implies that for larger molecules, which are likely to require 
deeper circuits with more two-qubit gates, increasingly higher two-qubit gate 
fidelities will be required to keep VQE viable for ground state computation.
This result should hold beyond the individual molecules and VQE algorithms discussed here.

\bibliography{bibliography}{}

\clearpage
\section*{Figure Legends}
Supplementary Fig. 1. Energies calculated using (solid lines) noiselessly-optimized and (dotted lines) noisily-optimized parameters, using the same ansätze.

Supplementary Fig. 2. Energies calculated using (solid lines) noiselessly-grown and (dotted lines) noisily-grown ansätze, with parameters optimized using gradients estimated with the finite difference method.

Supplementary Fig. 3. Energy error for $\mathrm{H}_4$, simulated using the fermionic-ADAPT-VQE, qubit-ADAPT-VQE and QEB-ADAPT-VQE (with the gradient decision rule). The solid lines are energies evaluated using noise applied after each ansatz element (the element-by-element, or EBE, method), while the dotted lines are energies evaluated using noise applied after each CNOT gate (the gate-by-gate, or GBG, method). The right figure uses a realistic noise probability of 0.1\%, whereas the left figure uses a noise probability representative of the values used in our simulations.

Supplementary Fig. 4. Noisy circuit diagrams of single qubit evolutions for each of the noise models used: (a) element by element noise model; (b) gate by gate noise model.

Supplementary Fig. 5. Energy accuracy as a function of gate-error probability for $\mathrm{H}_{2}$, $\mathrm{H}_4$ (at 1\AA\ and 3\AA\ interatomic separation), $\mathrm{LiH}$ and $\mathrm{HF}$. Energy accuracies lower than chemical accuracy are highlighted by the yellow region. The orange lines are the energies calculated using the k-UpCCGSD ansatz, for $k=1$, $k=2$ and $k=3$ \cite{k-UpCCGSD}. The red line is the fermionic-ADAPT-VQE ansatz, discussed in the main text. All ansätze shown use efficient fermionic elements.

Supplementary Fig. 6. Colour plots representing the energy error (from FCI) at different noise probabilities and ansatz lengths, for three different molecules. Ansätze are grown using the two different iterative-growth methods excluded from Fig.~3: fermionic-ADAPT-VQE (with inefficient elements) and QEB-ADAPT-VQE (with the gradient-based decision rule). The yellow line on the colour plot highlights the ansatz length which minimizes energy error for each noise probability.